\documentclass[twocolumn,english,aps,prd,floatfix,amssymb,superscriptaddress,longbibliography]{revtex4}
\usepackage[latin9]{inputenc}
\usepackage{verbatim}
\usepackage{float}
\usepackage{amsmath}
\usepackage{mathtools}
\usepackage{amssymb}
\usepackage{graphicx}
\usepackage{color}
\usepackage{xcolor}
\usepackage{tikz}
\usepackage[normalem]{ulem}
\newcommand{\ket}[1]{\ensuremath{\left| #1 \right>}}

\usepackage{bbold}

\usepackage[bookmarks=false,linkcolor=blue,urlcolor=blue,colorlinks,citecolor=blue]{hyperref}
\makeatletter

\newcommand{\be}{\begin{equation}}
\newcommand{\ee}{\end{equation}}
\newcommand{\bea}{\begin{eqnarray}}
\newcommand{\eea}{\end{eqnarray}}

\begin{document}

\title{Interplay between Symmetry Breaking and Interactions in a Symmetry Protected Topological Phase}

\author{Parameshwar R. Pasnoori}
\affiliation{Department of Physics, University of Maryland, College Park, MD 20742, United
States of America}
\affiliation{Laboratory for Physical Sciences, 8050 Greenmead Dr, College Park, MD 20740,
United States of America}

\author{Patrick Azaria}
\affiliation{Laboratoire de Physique Th\'eorique de la Mati\`ere Condens\'ee, Sorbonne Universit\'e and CNRS, 4 Place Jussieu, 75252 Paris, France}

\email{pparmesh@umd.edu}
\begin{abstract}

  We solve the one dimensional massive Thirring  model or equivalently the sine-Gordon model in the repulsive regime with general Dirichlet boundary conditions, which are characterized by two boundary fields $\phi_{L,R}$ associated with the left and right boundaries respectively. In the presence of these boundary fields, which explicitly break the charge conjugation symmetry, the system exhibits a duality symmetry which changes the sign of the mass parameter $m_0$ and shifts the values of the boundary fields by $\phi_{L,R}\rightarrow \phi_{L,R}+\pi$. When the mass parameter $m_0<0$ and the boundary fields $\phi_{L,R}=0$, or equivalently due to duality symmetry, when the mass parameter $m_0>0$ and the boundary fields $\phi_{L,R}=\pi$, the system is at a trivial point. Here, the ground state is unique just as in the case of periodic boundary conditions. In contrast, when the mass parameter $m_0<0$ and the boundary fields $\phi_{L,R}=\pi$, and equivalently due to the duality symmetry, when the mass parameter $m_0>0$ and the boundary fields $\phi_{L,R}=0$, the system is at a topological point where it exhibits a symmetry protected topological (SPT) phase, which is characterized by the existence of zero energy bound states at both the boundaries. As a result, the system exhibits two fold degenerate ground state in each fermionic parity sector and an emergent boundary Hilbert space. For a given value of the mass parameter $m_0$, we find that the bound state structure and the structure of the boundary Hilbert space remains intact even if the boundary fields take values way from the special values corresponding to the topological and trivial points, provided these values are less than certain critical values. We refer to these regions around the trivial and topological points as the trivial and mid-gap phases respectively. We find that the critical values depend on the strength of the interactions in the bulk, which is characterized by the Luttinger parameter $\beta$. Hence, we show that the stability of the SPT and trivial phases depends on the interplay of the symmetry breaking boundary field values and the bulk interaction strength.


\end{abstract}

\maketitle

\section{Introduction}

The hallmark of symmetry protected topological (SPT) phases is the robust ground state degeneracy which cannot be described  by the onset of a local bulk order parameter, but is rather due to a non local string order parameter \cite{Wen,Turner2011,Sau2011,Fidkowski2011b, Beenakker2013, ruhman2015, ruhman2017topological, jiang2017symmetry, kainaris2017interaction}. This ground state degeneracy is associated with the existence of zero energy modes (ZEM) localized at the ends of a system \cite{Starykh2000, Starykh2007,Ruhman2012,
 Zoller2013, Keselman2015, Diehl2015, chen2017flux, kainaris2017transmission, Scaffidi2017, Keselman2018}. Paradigmatic examples of SPT phases are the spin-1 Haldane chain \cite{HALDANE,AKLT} and one dimensional spin triplet charge conserving  superconductors \cite{Keselman2015,PAA1}, which are strongly interacting systems. The SPT phases can also arise in certain non interacting systems, the famous examples of which are the Kitaev chain \cite{Kitaev2001} and the Su-Schrieffer-Heeger (SSH) model \cite{ssh}. Usually, one applies simple open boundary conditions and as a result, the phase exhibited by the system is completely dictated by the bulk properties. For example, the one-dimensional superconductor  exhibits an SPT phase when the superconducting order parameter is spin-triplet. In contrast, its exhibits a trivial phase when the superconducting order parameter is spin-singlet \cite{Keselman2015,PAA1}. This is also the case with the SSH, which in the low energy is described by the one dimensional massive Dirac field theory. The system exhibits an SPT phase when the mass parameter is positive, whereas, it exhibits a trivial phase when the mass parameter is negative. When bulk interactions are turned on, the massive Dirac field theory which corresponds to the low energy limit of the SSH model, gives rise to the massive Thirring (MT) model \cite{Thacker}, or equivalently the sine-Gordon (SG) model \cite{MTSGcoleman}.
The SG model effectively describes low energy degrees of freedom of certain quantum circuits \cite{SGSPTcircuit,ROYSG}. It also describes the gapped spin degrees of freedom of the one dimensional superconductors mentioned above. On the other hand, the gapless charge degrees of freedom of these superconductors are described by the Luttinger liquid. Being massless, they are responsible for quasi-long range superconducting  correlations which decay as a power law. The Hamiltonian of the SG model is $H=\int_{0}^Ldx\; \mathcal{H}(x)$ where
\be
{\cal H}=\frac{1}{2} [(\partial_x \Phi(x))^2 + (\partial_x \Theta(x))^2] - \lambda \cos{\beta \Phi(x)},
\label{SG}
\ee
where $\Theta(x)$ is the dual of the $\Phi(x)$ field satisfying
$[\Theta(x), \Phi(y)]=i Y(x-y)$, where $Y(u)$ is the Heaviside function. When the mass parameter $\lambda >0$, the SG model (\ref{SG}) describes spin singlet superconductors (SSS), which are topologically trivial, whereas when $\lambda <0$ it describes topological  spin triplet superconductors (STS). 
The Luttinger parameter $\beta$ corresponds to the interaction strength. The system exhibits a mass gap for $0<\beta<\sqrt{8\pi}$, where $\beta\rightarrow 0$, corresponds to the semi-classical limit, and $\beta\rightarrow\sqrt{8\pi}$ corresponds to the weak coupling where the quantum fluctuations are maximal. There exists a special point $\beta=\sqrt{4\pi}$, where the SG model is equivalent to the free massive Dirac field theory discussed above. Although both models with $\lambda >0$ and $\lambda <0$
share the same bulk physics, they differ by their edge properties as mentioned above. To see this, consider imposing the boundary conditions
\be
  \Phi(0)=-\frac{\phi_L}{2\sqrt{\pi}}, \;  \Phi(L)=\frac{\phi_R}{2\sqrt{\pi}}
 \label{boundarySG}
\ee
on Eq. (\ref{SG}).
In the case $\Phi(0)=\Phi(L) = {0 \mod 2\pi/\beta}$, the SG model is invariant under the charge conjugation transformation, to be discussed below. In this case of zero boundary fields, when $\lambda<0$, the system was shown to host fractional solitonic charges (or equivalently spin 1/4) at the free fermion point $\beta=\sqrt{4\pi}$ \cite{JackiwRebbi,Jackiw}. Later it was shown  \cite{Keselman2015} using semi-classical arguments and DMRG techniques that in the $U(1)$ Thirring model, whose spin degrees of freedom are described by the SG model (\ref{SG}), these fractional spins survive interactions in the semi-classical limit $\beta \rightarrow 0$. It was also shown that the system exhibits an SPT phase which is protected by the $\mathbb{Z}_2$ symmetry corresponding to the charge conjugation mentioned above, where the system gives rise to ZEM. In \cite{PAA1}, the $U(1)$ Thirring model was solved exactly using Bethe ansatz, and it was shown that these ZEMs survive quantum fluctuations in the weak-coupling limit $\beta \rightarrow \sqrt{8\pi}$. In addition, it was argued that when the fractional spin 1/4 at the edges correspond to genuine quantum numbers, the system exhibits two Majorana zero modes (MZMs) at each edge. Recently, the SG model (\ref{SG}) was solved exactly using coordinate Bethe ansatz and it was shown to exhibit SPT phase for all values of $\beta$ and $\lambda$ \cite{MTpaper25}. 

In the presence of arbitrary boundary fields $\phi_L,\phi_R\ \neq 0$ in (\ref{boundarySG}), the charge conjugation symmetry is explicitly broken. There exist two types of boundary conditions: regular open boundary condition (OBC) or twisted open boundary conditions $\widehat{\text{OBC}}$, to be discussed below. In the limit $\beta\rightarrow \sqrt{8\pi}$, the $U(1)$ Thirring model was shown \cite{PAA2} to exhibit a duality symmetry between two pairs of bulk superconducting order parameter and boundary conditions: SSS-$\widehat{\text{OBC}}$ and STS-OBC correspond to a  topological boundary fixed point where the system exhibits an SPT phase with two fold degenerate ground state in each fermionic parity sector, while the pairs SSS-OBC and STS-$\widehat{\text{OBC}}$ correspond to a trivial fixed point where the ground state is unique. It was further shown that when the boundary fields take values away from that corresponding to the topological boundary fixed point, the SPT phase is stable provided these boundary values lie within a certain range. This naturally leads to the question whether the SPT phase is stable for non zero values of the boundary fields when $\beta$ takes values away from the weak coupling limit. To answer this we consider the SG model (\ref{SG}) with  generic integrable boundary conditions (\ref{boundarySG}). Using the bosonization formula
\be
\displaystyle{\Psi_{L,R}(x)= \frac{1}{\sqrt{2 \pi a_0} }e^{\mp i \sqrt{\pi}(\Phi(x)\pm \Theta(x))}},
\label{boso}
\ee
where $a_0$ is a short distance cutoff, one obtains the MT Hamiltonian density \cite{MTSGcoleman}
\bea
\nonumber\mathcal{H}=-i( \Psi^{\dagger}_R(x)\partial_x \Psi_R(x)-\Psi^{\dagger}_L(x)\partial_x\Psi_L(x))\\\nonumber+im_0\left(\Psi^{\dagger}_L(x)\Psi_R(x)-\Psi^{\dagger}_R(x)\Psi_L(x)\right)\\+2g\;\Psi^{\dagger}_R(x)\Psi^{\dagger}_L(x)\Psi_L(x)\Psi_R(x)\label{MThamiltonian},\eea
where $m_0=-\pi a_0 \lambda$ and $\beta^2/4\pi=2-2u/\pi$, $u=\frac{\pi}{2}-\text{tan}^{-1}\frac{g}{2}$.  Since the SG model describes SSS and STS phases when $\lambda >0$ and $\lambda < 0$ respectively, the MT model describes SSS and STS phases when $m_0 <0$ and $m_0 >0$ respectively. When $g=0$ the MT model corresponds to the free massive Dirac field theory as mentioned above. When $g>0$, the interactions in the system are repulsive, whereas, when
$g < 0$ the interactions in the system are attractive. These regimes correspond to the attractive $ \beta^2/4\pi<1$ and repulsive $ \beta^2/4\pi>1$ regimes of the SG model respectively. The model (\ref{MThamiltonian}) has been analyzed extensively \cite{MTMueller,MTLowenstein,MTWeisz,MTSGcoleman,MTHAHLEN,MtSgUhlenbrock} and has been solved using Bethe ansatz \cite{Thacker} with periodic boundary conditions. For the case of open boundary conditions, the massive Thirring model (\ref{MThamiltonian}) was analyzed using Bethe ansatz in the attractive regime in \cite{skorik}, where the boundary bound state structure was investigated. As mentioned before, it was recently analyzed in \cite{MTpaper25} for special values of the boundary fields, where the topological properties of the model were investigated.
  However, to the best of our knowledge, the solution for non zero values of the boundary fields in the repulsive regime, and the associated topological properties have not been obtained before. To this end, we solve the model with general open boundary conditions, to be provided below, using Bethe ansatz in the repulsive regime. We show that similar to the weak-coupling regime \cite{PAA2}, the model exhibits bulk-boundary duality symmetry. In the current model the duality symmetry changes the sign of the mass parameter $m_0$ and shifts the values of the boundary fields by $\phi_{L,R}\rightarrow \phi_{L,R}+\pi$. 
For a given mass parameter $m_0<0$, we find that when the values of the boundary parameters take values in a certain range $\phi_{L,R}\in (-u,u)$, or equivalently due to duality symmetry, when the boundary fields take values in the range $\phi_{L,R}\in (-\pi+u,\pi-u)$ \footnote{ $-\pi$ is identified with $\pi$, as $\phi\in (-\pi,\pi]$.} for mass parameter $m_0>0$, the system exhibits a trivial phase, where the ground state is unique and the system does not exhibit any boundary bound states. In contrast, for a given mass parameter $m_0>0$, when the values of the boundary parameters take values in a certain range $\phi_{L,R}\in (-u,u)$, or equivalently due to duality symmetry, when the boundary fields take values in the range $\phi_{L,R}\in (-\pi+u,\pi-u)$ for mass parameter $m_0<0$, the system exhibits mid-gap phase, where the system exhibits a bound state at both the boundaries. Within the mid-gap phase, when $\phi_{L,R}=0$ for $m_0>0$ or equivalently due to duality symmetry, when $\phi_{L,R}=\pi$ for $m_0<0$, which we refer to as the topological point, the system exhibits an SPT phase where the bound states at both the boundaries have zero energy, and the system exhibits a two fold degenerate ground state in each fermionic parity sector. Away from the topological point, these bound states at both the boundaries have finite energy which is less than bulk mass gap, and hence are referred to as mid-gap states. Even though the boundary bound states have finite energy, we find that the structure of the boundary Hilbert space throughout the mid-gap phase is same as that at the topological point. Hence, the stability of the SPT phase can be associated with the range of the boundary fields corresponding to the mid-gap phase. As mentioned above, the mid-gap phase exists when the values of the boundary fields lie within a certain range, where the magnitude of their values away from the topological point is less than the critical value $\phi_{C}=u$, which is related to the bulk interaction strength $g$, or equivalently the Luttinger parameter $\beta$. This critical value $\phi_C$ is maximum at the free fermion point $\beta=\sqrt{4\pi}$ and goes to zero in the weak coupling limit $\beta\rightarrow \sqrt{8\pi}$. Hence, the SPT phase is maximally stable at the Luther-Emery line, which corresponds to the free fermion limit as mentioned above, and as we move away from this free fermion point by turning on the interactions, the stability of the SPT phase with respect to perturbing the boundary fields decreases. We also find that the stability of the trivial phase also decreases with the increase in the bulk interaction strength in a similar manner. 

The paper is organized as follows. We start by discussing the symmetries and properties of the model in section (\ref{sec:symmetries}). We provide the exact wavefunction and the associated Bethe equations in section (\ref{sec:Betheansatz}). We discuss results in the trivial and SPT phases in sections (\ref{sec:trivial}), and (\ref{sec:topological}) respectively. In section (\ref{sec:phaseboundaries}), we discuss the stability of both the phases. In section (\ref{sec:discussions}), we provide the summary and discuss the future directions. The details of the Bethe ansatz calculations are provided in the appendices (\ref{apA}) and (\ref{apB}).

\section{Symmetries and Properties}
\label{sec:symmetries}
Before going further in our analysis let us recall some of the basic properties of the MT model and discuss the boundary conditions we shall consider. 
\subsection{Periodic boundary conditions}
The Hamiltonian (\ref{MThamiltonian}) is invariant under the $U(1)$ symmetry which conserves the charge 
\bea \label{tcharge} N=\int_0^L \Psi^{\dagger}_L(x)\Psi_L(x)+\Psi^{\dagger}_R(x)\Psi_R(x).\eea 
 Notice that unlike the $U(1)$ Thirring model \cite{Japaridze}, the chiral symmetry $\Psi_{R,L}(x)\rightarrow \Psi_{L,R}(x)$ is explicitly broken due to the presence of the mass term.  It is also invariant under the charge conjugation 
 \be \label{cc}\mathcal{C}\Psi^{\dagger}_{L,R}(x)=\Psi_{L,R}(x),\ee which has the $\mathbb{Z}_2$ group structure $\{1,\mathcal{C}\}$, $\mathcal{C}^2=1$. We will see later that the general open boundary conditions will explicitly break the charge conjugation symmetry.

The fundamental excitations in the system are solitons which have the same mass $m$ in both the SSS and STS phases. It is given by \cite{Thacker}

\begin{align}
m= \frac{|m_0|\gamma}{\pi(\gamma-1)}\tan(\pi\gamma)\;e^{\Lambda(1-\gamma)}, \label{mass} \\\text{where}\;\;\; \gamma=\frac{\pi}{2u}, \;\; u=\frac{\pi}{2}-\text{tan}^{-1}\frac{g}{2}, \nonumber
\end{align}
where $u$ is a renormalized coupling and $\Lambda=\log{(D/|m_0|)}$ where $D$ is an UV cutoff. A restriction $u > \pi/3$ is imposed by the regularization scheme  \cite{Thacker}. Contact with the SG model is made through the scaling argument 
\be
\frac{m}{D} \propto (\frac{|m_0|}{D})^{\frac{1}{2-\beta^2/4\pi}}
\ee
which yields $\beta^2/4\pi=2-2u/\pi$ which matches with Coleman's expression  for small $g$: i.e: $\beta^2/4\pi \simeq 1+g/\pi$.  
In the scaling limit the dimensionless bare mass term $|m_0|/D$ goes to zero as the cutoff $D$ is sent to infinity while keeping the renormalized mass m fixed, i.e: $|m_0|/D \propto (m/D)^{2u/\pi}$. Let us finally note that in terms of the renormalized coupling $u$ in the MT model, the free fermion point is obtained with $u=\pi/2$, while attractive interactions are described by $u >\pi/2$ and repulsive interactions are described by $u <\pi/2$. 


\subsection{General open boundary conditions}
We now consider the  general, charge conserving,  open boundary conditions $\text{OBC}$
\bea\Psi_L(0)=e^{i\phi}\Psi_R(0), \;\;\Psi_L(L)=-e^{ -i\phi^{\prime}}\Psi_R(L).\label{bcMT}\eea
We chose the boundary phases $\phi$ and $\phi^{\prime}$ to depend on the renormalized coupling $u$ as follows
\bea\text{OBC}:&& \phi=\phi_L+u-\pi/2, \;\; \phi^{\prime}=\phi_R+u-\pi/2
\label{anomalousBC}
\eea
where $\phi_L$ and $\phi_R$ correspond to the  sine-Gordon boundary fields given in (\ref{boundarySG}).
This extra term 
$u-\pi/2$ in (\ref{anomalousBC}), arises due to boundary anomaly \cite{skorik}, and is
necessary for the solution of 
of the massive Thirring model to be consistent with that of the sine-Gordon model. The boundary conditions (\ref{bcMT}) conserve the total charge (\ref{tcharge}) but break the charge conjugation symmetry (\ref{cc}). However, as discussed in the Appendix (\ref{apA}), the Bethe equations are invariant under the combined application of charge conjugation $\mathcal{C}$ and the change of  signs of the boundary fields $\phi_{L,R}\rightarrow-\phi_{L,R}$. 


\begin{figure}[h!]
\includegraphics[width=0.8\columnwidth]{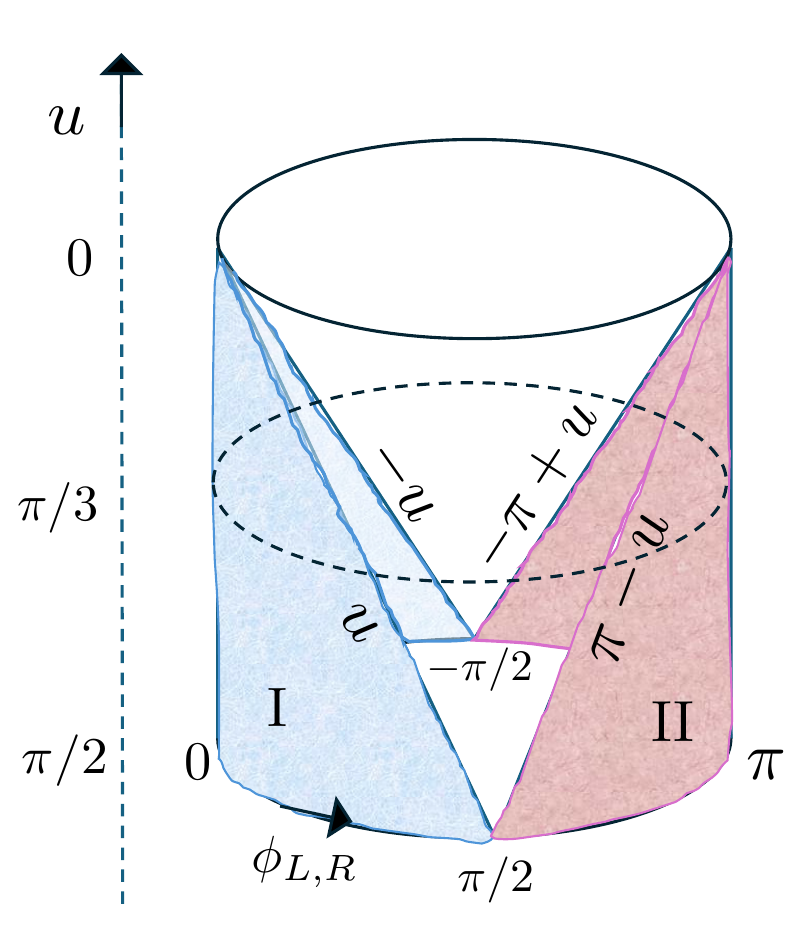}
\caption{The figure shows different ranges of the values of the boundary fields $\phi_{L,R}$ in the OBC (\ref{bcMT}). There exists two regions I (blue) and II (red) corresponding to the following values of the boundary fields: I ($-u<\phi_{L,R}<u$), II ($-\pi<\phi_{L,R}<-(\pi-u)$ and $\pi-u<\phi_{L,R}<\pi$), where $-\pi$ is identified with $\pi$. For $m_0>0$, the system exhibits a mid-gap phase when $\phi_{L,R}\in \text{I}$, whereas it exhibits trivial phase when $\phi_{L,R}\in \text{II}$. Due to the duality symmetry (\ref{duality}), in addition to change in the sign of $m_0$, the range of the values of $\phi_{L,R}$ corresponding to the regions I and II described above get interchanged. Hence, for $m_0>0$, the  region I (blue) corresponds to the trivial phase whereas the region II (red) corresponds to the mid-gap phase. Note that at the Luther-Emery line, the boundaries of the mid-gap and trivial phases intersect at $\phi_{L,R}=\pm\pi/2$. As the repulsive interaction strength is turned on, the range of the fields $\phi_{L,R}$ decreases and eventually shrinks to zero when approaching the weak coupling limit $u\rightarrow 0$. }  \label{pdrange}
\end{figure}

\subsection{Duality Symmetry}

 The system exhibits a duality symmetry $\Omega$ 
\bea \Omega\Psi(x)&=&\hat{\Psi}(x)\\
\hat{\Psi}_L(x)=\Psi_L(x), &&\hat{\Psi}_R(x)=-\Psi_R(x).\label{duality}\eea
In a system with PBC, the duality $\Omega$ relates  models with opposite values of the bare masses $m_0$ and the same $g$:
$ \mathcal{H}(\Psi,g,m_0)=\mathcal{H}(\hat{\Psi},g,-m_0)$. In the presence of general OBC (\ref{anomalousBC}) the duality also changes the boundary conditions to dual ones, i.e:
$\phi \rightarrow \phi + \pi$ and 
$\phi^{\prime} \rightarrow \phi^{\prime} + \pi$ that we shall denote
\bea\widehat{\text{OBC}}:&& \phi=\phi_L+u+\pi/2, \;\; \phi^{\prime}=\phi_R+u+\pi/2.
\label{anomalousdualBC}
\eea
Hence the duality symmetry $\Omega$ provides for an isometry that relates, at all energy scales, the system  in  the STS phase with $m_0>0$ and OBC boundary conditions (\ref{anomalousBC}) to the system in the SSS phase with $m_0<0$ and $\widehat{\text{OBC}}$ boundary conditions (\ref{anomalousdualBC}) and vice versa,i.e:
%
\bea \mathcal{H}(\Psi,m_0,\text{OBC})=\mathcal{H}(\hat{\Psi},-m_0,\widehat{\text{OBC}}). 
\eea
As we shall see, the above duality will play an important role in the boundary physics.

\section{Bethe Ansatz equations}
\label{sec:Betheansatz}
Since the number of particles $N$ (\ref{tcharge}) is a good quantum number, one can construct the Bethe wave function labelled by $N$ as follows
\begin{align}
\nonumber \ket{N} =    \sum_{\alpha_1,\dots,\alpha_N=L,R}\int &dx_1 \dots dx_N \mathcal{A}\Upsilon^{\alpha_1...\alpha_N}_{\beta_1...\beta_N}(x_1,...x_N) \\
&\Psi^{\dagger}_{\alpha_1}(x_1)\dots \Psi^{\dagger}_{\alpha_N}(x_N)\ket{\text{vac}}.
\end{align}
Here  the $\beta_i$ correspond to the rapidities, $\mathcal{A}$ represents anti-symmetrization with respect to the indices $\alpha_i,x_i$ and 
\begin{align} \nonumber 
& \Upsilon^{\alpha_1...\alpha_N}_{\beta_1...\beta_N}(x_1,...x_N) \nonumber \\
& = \sum_Q \theta(\{x_{Q(j)}\}) \smashoperator{\sum_{\delta_1,\dots,\delta_N=1,2}} A_Q^{\delta_1...\delta_N} \prod_{i=1}^{N} Y^{\alpha_i}_{\delta_i,\beta_i}(x_i).
\end{align}
In this expression, $Q$ denotes a permutation of the position orderings of particles and  $\theta(\{x_{Q(j)}\})$ is the Heaviside function that vanishes unless $x_{Q(1)} \le \dots \le x_{Q(N)}$. $A^{\delta_1...\delta_N}_Q$ denote amplitudes corresponding to the respective ordering of the particles, and 
\begin{align}
\nonumber &Y_{1\beta}^R(x) =-ie^{-\beta/2}e^{-im_0x\sinh\beta}, \\
&Y_{2\beta}^R=ie^{\beta/2}e^{im_0x\sinh\beta},\nonumber \\ 
&Y_{1\beta}^L(x)=e^{\beta/2}e^{-im_0x\sinh\beta}, \text{and} \nonumber \\
&Y_{2\beta}^L=-e^{-\beta/2}e^{im_0x\sinh\beta}.\label{NPWF}
\end{align}
Applying the Hamiltonian (\ref{MThamiltonian}) to the $N$-particle wavefunction (\ref{NPWF}), and imposing the boundary conditions (\ref{bcMT}), we obtain the following set of constraint equations for the rapidities $\beta_j$, called the Bethe equations:

\begin{widetext}
\bea \nonumber e^{2im_0L\sinh\beta_i}=\frac{\cosh\left(\frac{1}{2}(\beta_i+i\phi_L+iu)\right)}{\cosh\left(\frac{1}{2}(\beta_i-i\phi_L-iu)\right)}\frac{\cosh\left(\frac{1}{2}(\beta_i+i\phi_R+iu)\right)}{\cosh\left(\frac{1}{2}(\beta_i-i\phi_R-iu)\right)}\prod_{j\neq i,j=1}^N\frac{\sinh\left(\frac{1}{2}(\beta_i-\beta_j+2iu)\right)}{\sinh\left(\frac{1}{2}(\beta_i-\beta_j-2iu)\right)}\frac{\sinh\left(\frac{1}{2}(\beta_i+\beta_j+2iu)\right)}{\sinh\left(\frac{1}{2}(\beta_i+\beta_j-2iu)\right)}.\\\label{BE}\eea
\end{widetext}

Each eigenstate of the Hamiltonian corresponds to a unique set of roots $\{\beta_j\}$ which satisfy  (\ref{BE}). Due to the presence of open boundary conditions, the Bethe equations are  symmetric under the reflection $\beta_i\leftrightarrow -\beta_i$, so that the solutions to (\ref{BE}) come in pairs $(\beta_i,-\beta_i)$. In the thermodynamic limit $N,L\rightarrow\infty$, the roots form a dense set, and the Bethe equations can be expressed as integral equations which can be solved using a Fourier transform. The energy of the thus obtained eigenstate is given by
\bea E=\sum_{j=1}^N m_0\cosh(\beta_j).\label{eneq}
\eea
In the following, assuming the OBC (\ref{bcMT}), we shall discuss the solutions to the Bethe equations (\ref{BE}) in the two domains of the boundary fields
\bea \nonumber
\text{I} &:& \;\; -u< \phi_{L,R} <u,\\\nonumber
\text{II}&:& -\pi<\phi_{L,R}<-(\pi-u) , \;\;\;\pi-u<\phi_{L,R}<\pi,\\
\label{region}
\eea
which are depicted in Fig. \ref{pdrange}. Under the duality transformation (\ref{duality}), the boundary fields in region I and II are mapped onto each other. For a given sign of the mass parameter $m_0$, the system exhibits both trivial and mid-gap phases depending on the values of the boundary fields $\phi_{L,R}$ as shown in Fig. \ref{pdrange}.  For $m_0<0$, region I corresponds to the trivial phase and region II corresponds to the mid-gap phase. The duality symmetry (\ref{duality}) changes the sign of $m_0$ and exchanges the values of the boundary fields in regions I and II.  Thus, for $m_0>0$, region II corresponds to the trivial phase and region I corresponds to the mid-gap phase. As we shall show below, in the trivial phase the system exhibits a unique ground state with no excited states below the soliton mass (\ref{mass}).  In the mid-gap phase, for general values of the boundary fields, there are four states with energies less than twice the solition mass (\ref{mass}). One among them is the ground state and the rest of the states contain bound states exponentially localized at the boundaries. These bound states are associated with special solutions of Bethe equations called ``short'' or ``close'' boundary strings \cite{PAA1,PAA2,Parmeshthesis, ParmeshKondo1,YSRKondo,XXXphases,susypaper}. In the uncolored regions of Fig. \ref{pdrange}, the Bethe equations have a different kind of boundary string called ``wide'' boundary strings \cite{YSRKondo,Parmeshthesis,XXZterras} which describe a rather complicated boundary phenomena unlike the simple exponentially localized bound states. In the following, we shall discuss the Bethe ansatz results in the trivial and mid-gap phases, relegating the details of the Bethe ansatz calculation to the appendices.

\section{Trivial Phase}
\label{sec:trivial}
We start by describing the trivial phase.  We discuss the case $m_0 <0$ with boundary fields  lying in region I (\ref{region}). Due to the duality symmetry, this is equivalent to the case $m_0>0$ with boundary fields lying in the region II as mentioned above. In the trivial phase, the Hamiltonian has  a unique ground state. One obtains its normal ordered charge by  removing from (\ref{tcharge}) the bulk contribution which is the charge associated with  a system with periodic boundary conditions (see the Appendix (\ref{apB}) for details).  We find that the normal ordered charge of the ground state is
\be
:N:  = q \left(\frac{\phi_L + \phi_R}{2 \pi}\right),
\label{GSchargetrivial}
\ee
where 
\be
q=\frac{\pi}{2(\pi -u)}
\label{solcharge}
\ee
is the  soliton charge. Note that the contribution to the normal ordered charge is solely due to the boundary fields. Hence, the ground state charge (\ref{tcharge}), when expressed in units of the soliton charge  ${\cal N} =:N:/q$, identifies with the topological charge of the SG model. Indeed, (\ref{boundarySG}) and (\ref{GSchargetrivial}) together imply that
\be
{\cal N} = \frac{1}{\sqrt \pi} \int_0^L dx\; \partial_x \Phi(x).
\label{SGcharge}
\ee
One could have defined the charge operator by reabsorbing the boundary field contribution of the ground state (\ref{GSchargetrivial}):
\be
{\cal N} \rightarrow {\cal N} - \left(\frac{\phi_L + \phi_R}{2 \pi}\right).
\label{normalcharge}
\ee
This corresponds to another normal ordering prescription in which the charge of all states are integers. According to this definition, the ground state has charge ${\cal N}= 0$. This allows us to label the ground state in the trivial phase as

\bea \ket{GS}=\ket{0}.
 \label{GStrivial}
 \eea

Note that at the trivial point, which corresponds to $\phi_L=\phi_R=0$ for $m_0<0$, both the normal ordering prescriptions coincide. With this new normal ordering prescription, one can define the fermionic parity \bea \mathcal{P}= (-1)^{\mathcal{N}}  \label{fermionicparity} \eea associated with each state. We can see that the ground state in the trivial phase has even fermionic parity.

The fundamental  excitation on top of the ground state is either a soliton or an anti-soliton, and they carry charge ${\cal N}= 1$ or ${\cal N}= -1$ respectively. They both have energy
\be m_{\theta}=m\cosh\theta,
\ee
where $\theta$ is the rapidity and $m$ is the mass gap (\ref{mass}). Note that when the total charge of the system is fixed, one is only allowed to add pairs of solitons and anti-solitons. One of the key signatures of the trivial phase is that all the excited states on top of the unique ground state contain solitons and anti-solitons which form continuum of excited states.

\section{Mid-Gap Phase}
\label{sec:topological}

\begin{figure}[h]
\includegraphics[width=0.8\columnwidth]{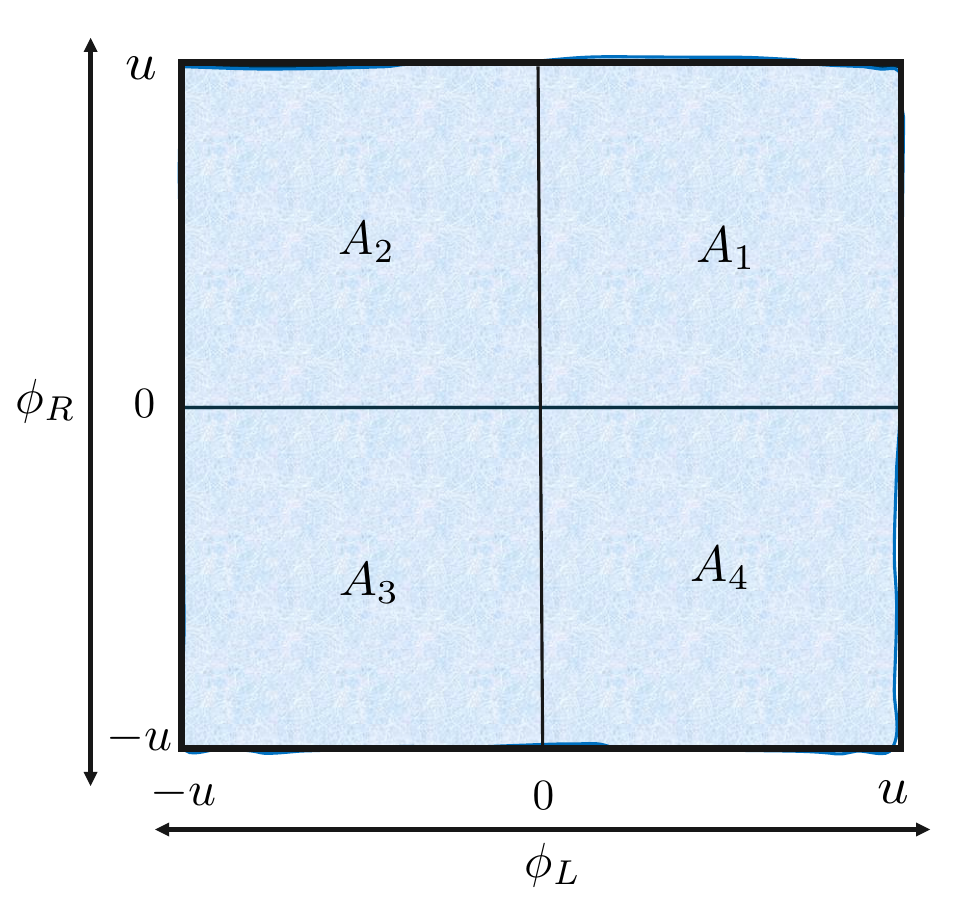}
\caption{The figure shows the boundary phase diagram exhibited by the system in the mid-gap phase. The x-axis and y-axis correspond to the boundary fields at the left and the right boundaries respectively. The system exhibits  four $A$ phases, which host bound states at both the edges. At the origin $\phi_L=\phi_R=0$, which corresponds to the topological point, both the boundary bound states carry zero energy which gives rise to two fold degenerate ground state in both the even and odd charge sectors. Away from the origin where both the boundary fields $\phi_L,\phi_R$ take values $\phi_L,\phi_R \in (-u,u)$, the boundary bound states carry finite energy less than the mass gap. As a result, the degeneracy of the ground states is lifted and all but one of the ground states become mid-gap states. } \label{pd}
\end{figure}

Now let us consider the mid-gap phase, where $m_0>0$ with the boundary fields lying in region I. Due to the duality symmetry, this is equivalent to the boundary fields lying in the region II with $m_0<0$ as mentioned above. In contrast to the trivial phase, there exist four low energy states which do not belong to the continuum.  We label these states according to their charges ${\cal N}$, as defined in (\ref{normalcharge}), writing 
\be
\ket{-1}, \ket{0}_{\cal L}, \ket{0}_{\cal R}, \ket{+1}. \label{Astates} 
\ee
We can see that the states $\ket{\pm 1}$ correspond to the odd fermionic parity sector, whereas, the states $\ket{0}_{\cal L}$ and $\ket{0}_{\cal R}$ correspond to even fermionic parity sector. They have the following energies with respect to that of the $|-1 \rangle$ state
  \bea
E_{|0\rangle_{\cal L,R}}&=& E_{  |-1 \rangle}+ m_{\cal{L,R}} \nonumber \\
E_{|+1\rangle}&=& E_{  |-1 \rangle}+ m_{\cal L} + m_{\cal R},
\label{MSenergies}
  \eea
  where
  \be\label{bsenergy} m_{\cal{L,R}}=m\sin(\gamma \phi_{L,R}),\; \gamma=\pi/2u
\ee
and $m$ is the soliton mass (\ref{mass}).
Physically, (\ref{bsenergy}) gives the energies of bound states are that are exponentially localized at the left and right edges of the system. The states $|0\rangle_{\cal{L,R}}$ are obtained from $ |-1 \rangle$
by adding a bound state at the left or right edge respectively
whereas the state $ |+1 \rangle$
is obtained by adding bound states at both edges. For each bound state added, the charge (\ref{normalcharge}) increases by 1.   From Eq. (\ref{bsenergy}), we see that the bound state energies may take values $-m \le m_{\cal{L,R}} \le m$, including negative values. As a result, the signs of $\phi_{L,R}$ determine which of the states (\ref{Astates}) has the lowest energy. As shown in Fig. \ref{pd} we  distinguish among four sub-phases labelled $A_1,A_2,A_3,A_4$, which correspond to the ranges of the boundaries fields given in Table \ref{table0}.
\begin{table}[h!]
\centering
\caption{Sub-phases and their properties}
\begin{tabular}{|c|c|c|c|}
\hline
\hline
Sub-phase  & $\phi_L$ Range & $\phi_R$ Range & Lowest energy\\
& & & state\\
\hline
$A_1$ & $0<\phi_L<u$ & $0<\phi_R<u$ & $|-1 \rangle$  \\
$A_2$ & $-u<\phi_L<0$  & $0<\phi_R<u$   & $|0\rangle_{\cal L}$ \\
$A_3$ & $-u<\phi_L<0$  & $-u<\phi_R<0$  & $|+1\rangle$ \\
$A_4$ & $0<\phi_L<u$   & $-u<\phi_R<0$  & $|0\rangle_{\cal R}$ \\
\hline
\hline
\end{tabular}
\label{table0}
\end{table}
The table also indicates the lowest energy state in each sub-phase.  Since  $|m_{\cal{L,R}}| \le m$, the remaining 3 states in (\ref{Astates}) are mid-gap states with energies below the soliton mass gap $m$ (\ref{mass}).  At the borders of region I when $|\phi_{L,R}| \rightarrow \phi_C=u$, their energies reach up to the mass gap, merging with the continuum of soliton states.

Before going further, let us pause and explain how the bound state picture described above emerges from the Bethe ansatz calculation.  We focus on sub-phase $A_1$ for concreteness. In the Bethe ansatz approach, in sub-phase $A_1$, $|-1 \rangle$ is obtained by choosing all Bethe roots to lie on the $i\pi$ line. Then, $|1\rangle$ is defined by applying charge conjugation to $\ket{-1}$. 
The two states  $ |0\rangle_{\cal L}$ and $ |0\rangle_{\cal R}$ are obtained by adding boundary string solutions of the Bethe equations to the high energy state $|1\rangle$.  These purely imaginary solutions have rapidities  $\beta_{R}= \pm i (u-\phi_R)+i\pi$  and $\beta_{L}= \pm i (u-\phi_L)+i\pi$ respectively. 
By adding $\beta_{R}$ to $|1\rangle$, we remove the right bound state and obtain $|0\rangle_{\cal L}$.  Similarly, by adding $\beta_{L}$ to $|1\rangle$, we remove the left bound state and obtain $|0\rangle_{\cal R}$. 
Under charge conjugation (\ref{cc}), $\ket{0}_{\cal{L}}\leftrightarrow \ket{0}_{\cal{R}}$. Analogous constructions apply to the sub-phases $A_2,A_3$, and $A_4$.
One can define the local boundary bound state parities associated with each state:
\bea \mathcal{P}_{\cal L}= (-1)^{n_{\cal L}},\; 
\mathcal{P}_{\cal R}= (-1)^{n_{\cal R}},
\label{bsparity}
\eea
where $n_{\cal L,R}=0,1$ is the number of boundary bound states at the left and right edges respectively. This allows us to label the four mid-gap states described above using the local boundary bound state parities as shown in Table \ref{table1}.  Note that the local bound state parities and the total fermionic parity are related by the following constraint

\be
\mathcal{P}=-\mathcal{P}_{\cal L}\mathcal{P}_{\cal R}.
\ee

\begin{table}[h!]
\centering
\caption{The ground state and the three mid-gap states and their respective local boundary bound state parities $\cal{P}_{\cal L,R}$. }

\begin{tabular}{|c|c|c|}
\hline
\hline
  State  & \;\;$\cal{P}_{\cal L}$ &\;\; $\cal{P}_{\cal R}$ \\
\hline
 $|-1 \rangle$& 1  & 1    \\
  $|0\rangle_{\cal{L}}$ & -1       &   1    \\
    $|0\rangle_{\cal{R}} $ & 1  & -1  \\
$|1\rangle$ & -1 & -1\\
\hline
\hline
\end{tabular}
\label{table1}
\end{table}  
In the remaining $A_2,A_3,A_4$ sub-phases,  the charge quantum numbers and their respective local boundary bound state parities are analogous to those in the $A_1$ sub-phase. The four sub-phases  $A_{j=1,...,4}$  transform into each other under charge conjugation (\ref{cc}) as $A_1 \leftrightarrow A_3$ and $A_2 \leftrightarrow A_4$. 

Now let us consider the topological point, which corresponds to $\phi_L=\phi_R=0$ for $m_0>0$. Here, the boundary conditions (\ref{bcMT}) as well as the Hamiltonian (\ref{MThamiltonian}) are invariant under charge conjugation (\ref{cc}). We see from (\ref{bsenergy}) that both the left and right bound states have zero energy. Due to this, if ${\cal N} = 0$, there exist two degenerate ground states $\ket{0}_{\cal L}$ and $\ket{0}_{\cal R}$. If ${\cal N} = 1$, there is a non-degenerate ground state $\ket{+1}$ and if ${\cal N} = -1$, there is a non-degenerate ground state $\ket{-1}$. Hence, there exist two degenerate states in both the even and odd fermionic parity sectors.

\section{Phase boundaries}
\label{sec:phaseboundaries}
In the mid-gap phase, the boundary bound states have energy (\ref{bsenergy}), which is always less than the mass gap $m$. When the absolute value of the boundary field at either the left or the right boundary approaches the critical value $\phi_C=u$, the energy of the bound state at the respective boundary approaches the mass gap $m_{\cal{L,R}}\rightarrow m$. As one increases the boundary field past the critical value, we move from region I ($m_0<0$) or region II ($m_0>0$) into the uncolored region in Fig \ref{pdrange}. As we cross these phase boundaries, as mentioned before, the close boundary string which describes the boundary bound state turns into a wide boundary string and the physics at the boundary is much more intricate. The localized bound state is no longer exponentially localized at the boundary and hence the above described mid-gap phase ceases to exist. Similarly, starting from the trivial phase, as we increase the values of the boundary fields past the critical values, we move from region I ($m_0>0$) or region II ($m_0<0$) into the uncolored region. This results in the emergence of wide boundary strings at the respective boundaries, and as a result, the trivial phase described in the previous section ceases to exist. 

The critical values of the boundary fields depend on the bulk interaction strength such that, as the bulk interaction strength is increased, the critical value decreases and goes from maximum at the Luther-Emergy line $\beta=\sqrt{4\pi}$ to zero at $\beta= \sqrt{8\pi}$. As the range of the values of the boundary fields corresponding to the mid-gap phase can be associated with the stability of the SPT phase, we see that both the SPT and the trivial phases are less stable with respect to perturbing the boundary fields as the interactions in the bulk get stronger, demonstrating that the stability of these phases depends on the interplay of the boundary fields and the bulk interaction strength.

\section{Summary and open questions}
\label{sec:discussions}

In this work we have analyzed the massive Thirring model with Dirichlet boundary conditions in the presence of boundary fields $\phi_{ L,R}$. We have obtained the Bethe equations and solved them  for repulsive interactions ($g \ge 0$) in two regions: region I, i.e: $\phi_{L,R}\in (-u,u)$, and region 
 II, i.e: $\phi_{ L,R}\in (-\pi+u,\pi-u)$. 
 The nature of our solution crucially depends on the sign of the bare fermion mass parameter $m_0$. For $m_0<0$
the system is shown to exhibit a trivial phase when the boundary fields lie in the region I whereas when they lie in the region II the system is in the mid-gap phase. Due to a duality symmetry, i.e: $m_0 \rightarrow -m_0$, $\phi_{ L,R}\rightarrow \phi_{ L,R}+\pi$, the nature of the two regions I and II  are interchanged when $m_0>0$.

 In the trivial phase, the system exhibits a unique ground state for all values of the boundary fields in the allowed range mentioned above. Within this phase, there exists a special point called the trivial point which corresponds to $\phi_{L,R}=0$ for $m_0<0$, or equivalently due to duality symmetry, it corresponds to $\phi_{L,R}=\pi$ when $m_0>0$, where the system exhibits charge conjugation symmetry. In the trivial phase, the values of the boundary fields is such that the magnitude of the difference from that at the trivial point is less than the critical value $\phi_C=u$.  In the mid-gap phase, the system exhibits boundary bound states at both the boundaries. These bound states have the same charge as that  of the soliton and their  energies  are  less than  the soliton mass. Due to this, above the ground state, the system exhibits three states whose maximum energy is below twice the  soliton mass. These   are referred to as mid gap states. At the special values of the boundary fields $\phi_{ L,R}=0$ for $m_0>0$ and $\phi_{ L,R}=\pi$ for $m_0<0$ the system is at the topological point, where the system exhibits an SPT phase protected by the $\mathbb{Z}_2$ symmetry associated with the charge conjugation. There the energies of the boundary bound states are  zero and, as a result, the three mid gap states become degenerate with the ground state resulting in a two fold degenerate ground state in each fermionic parity sector. In the mid-gap phase, the values of the boundary fields is such that the magnitude of the difference from that at the topological point is less than the critical value $\phi_C=u$.
This critical value decreases as the bulk interaction strength increases such that it is maximum at the Luther-Emery line $\beta=\sqrt{4\pi}$ and goes to zero in the weak coupling limit $\beta\rightarrow \sqrt{8\pi}$. Hence, we find that in the presence of the boundary fields which break the symmetry responsible for the existence of the SPT and trivial phases, the stability of these phases decreases as the bulk interaction strength increases, showing that the stability of these phases depends on the interplay between the boundary fields and the bulk interaction strength.

So far the results presented in this work are restricted to repulsive interactions and to the region I and II of boundary fields. For generic values of $\phi_{L,R}$ lying outside
these two regions a different type of boundary excitation as the one found in the mid-gap phase
are found. In the Bethe ansatz framework they are described by `wide boundary strings' \cite{PAA2} in contrast with the `closed boundary string' encountered in this work. The nature of the intermediate phase, as well as that of the (boundary) quantum phase transitions at the boundaries between  this phase and the mid gap and trivial ones are currently under investigation \cite{MTmodelpart3}. 
As well known, in the attractive regime on top of  solitons there exits   bulk  breathers  excitation. In a recent work \cite{MTpaper25}, it is found that in addition to the localized boundary bound states, there also exist boundary breathers whose mass scale is much smaller than that of the soliton resulting in a rich boundary excitation spectrum in the SPT phase. In the attractive regime when the boundary fields are present, the excitation spectrum and the resulting phase diagram is found to be much more intricate than in the repulsive regime. Due to the already lengthy size of the present paper we shall report our results for the attractive regime in a forth coming publication \cite{MTmodelpart2}.

\acknowledgements
We acknowledge helpful comments by A. Mizel. 

\bibliography{refMT}
\newpage
\appendix
\begin{widetext}

\section{The Bethe wave function}
\label{apA}

The Hamiltonian of the Massive Thirring model is
\bea\nonumber
H=-i\int_{0}^{L} \Psi^{\dagger}_R(x)\partial_x \Psi_R(x)+i\int_{0}^{L}\Psi^{\dagger}_L(x)\partial_x\Psi_L(x)+im_0\int_{0}^{L}\Psi^{\dagger}_L(x)\Psi_R(x)-\Psi^{\dagger}_R(x)\Psi_L(x)\\+2g\int_{0}^{L}\Psi^{\dagger}_R(x)\Psi^{\dagger}_L(x)\Psi_L(x)\Psi_R(x)\label{apMThamiltonian}.\eea

Here $\Psi_{L,R}(x)$ are the fermionic fields, $m_0$ is the bare mass and $g$ is the bulk interaction strength. We apply the following Dirchlet boundary conditions on the fermionic fields 
\bea
\Psi_L(0)=e^{i\phi}\Psi_R(0), \hspace{4mm} \Psi_L(L)=-e^{-i\phi^{\prime}}\Psi_R(L),\label{apbcMT1}
\eea
where $\phi,\phi^{\prime}$ are phases which are related to the physical boundary fields $\phi_L,\phi_R$ of the sine-Gordon model through the following relations 
\bea \label{apobc}\text{OBC}:\phi=\phi_L+u-\pi/2, \;\; \phi^{\prime}=\phi_R+u-\pi/2,\\
\eea
and  $u=\pi/2-\text{tan}^{-1}(g)$.
As discussed in the main text, the system exhibits a duality symmetry 
that relates the Hamiltonian with mass parameter $m_0$ and open boundary conditions (OBC) to that with mass $-m_0$ and dual open boundary conditions ($\widehat{\text{OBC}}$)
(\ref{apbcMT1}):
\bea \label{apduality}\mathcal{H}(\Psi,m_0,\text{OBC})=\mathcal{H}(\hat{\Psi},-m_0,\widehat{\text{OBC}}),
\eea
where 
\bea\label{aptobc}
\widehat{\text{OBC}}:\phi=\phi_L+u+\pi/2, \;\; \phi^{\prime}=\phi_R+u+\pi/2.
\eea
Since the duality transformation (\ref{apduality}) maps the system with $m_0<0\; (m_0>0)$ and OBC to the system with $m_0>0\;(m_0<0)$ and $\widehat{\text{OBC}}$, it suffices to solve the Hamiltonian (\ref{apMThamiltonian}) with $m_0>0$ with both OBC and $\widehat{\text{OBC}}$. Below we shall provide the details of the construction of the Bethe wave function and obtain the Bethe equations by applying the appropriate boundary conditions. In Appendix (\ref{apB}), we solve the Bethe equations, obtain the spectrum associated with the boundaries, and analyze the boundary phase diagram. Since the Hamiltonian (\ref{apMThamiltonian}) conserves the total number of particles $N$
\bea \label{aptcharge} N=\int_0^L \Psi^{\dagger}_L(x)\Psi_L(x)+\Psi^{\dagger}_R(x)\Psi_R(x),\eea 
we shall  construct the Bethe ansatz wave function in each  $N$ sector.

\subsection{One particle sector}

In one particle sector, the wave function can be written as
\bea
\ket {1}= \int_{0}^{L}dx\left(\chi^R_{\beta}(x)\Psi^{\dagger}_R(x)+\chi^{L}_{\beta}(x)\Psi^{\dagger}_L(x)\right)\ket{0},\label{1pwf}
\eea
where $\beta$ is the rapidity. Applying the Hamiltonian (\ref{apMThamiltonian}), to the one particle wave function (\ref{1pwf}), we obtain the following set of equations

\bea \nonumber(-i\partial_x -E) \chi_{\beta}^R(x)-im_0\chi_{\beta}^L=0,\\(i\partial_x -E) \chi_{\beta}^L(x)+im_0\chi_{\beta}^R=0,\label{1peq}\eea
where $E$ is the energy. To solve the above set of equations, we use the following ansatz for $\chi_{\beta}^{L,R}(x)$:
\begin{align}
\nonumber\chi_{\beta}^R(x) =\mathcal{N}\sum_{i=1}^2Y_{i\beta}^R(x) A_i(\beta,\phi) \;\theta(x)\theta(L-x),& \;\;\;\;\chi_{\beta}^L(x)=\mathcal{N}\sum_{i=1}^2Y_{i\beta}^L(x) A_i(\beta,\phi)\; \theta(x)\theta(L-x),\\\nonumber
Y_{1\beta}^R(x) =-ie^{-\beta/2}e^{-im_0x\sinh\beta}&,\;\;\; \;\;Y_{2\beta}^R=ie^{\beta/2}e^{im_0x\sinh\beta},\\ Y_{1\beta}^L(x)=e^{\beta/2}e^{-im_0x\sinh\beta}&,\;\;\;\;\;Y_{2\beta}^L=-e^{-\beta/2}e^{im_0x\sinh\beta}.\label{1pansatz}
\end{align}
Here $\theta(x)$ is the Heaviside function, with $\theta(x)=1$ for $x > 0$ and $\theta(0)=1/2$. $\mathcal{N}$ is a normalization constant and $A_i(\beta,\phi)$ are amplitudes which are constrained by the boundary conditions (\ref{apobc}),(\ref{aptobc}) as we shall see below. We find that the ansatz (\ref{1pansatz}) satisfies (\ref{1peq}), provided 
\bea E=m_0\cosh\beta.\eea

\vspace{3mm}

In addition, one obtains the following set of equations due to the presence of the boundaries: 

\bea \Psi^{\dagger}_R(0)\sum_iY_{i\beta}^R(0)A_i(\beta,\phi)-\Psi^{\dagger}_L(0)\sum_iY_{i\beta}^L(0)A_i(\beta,\phi)=0,
\\
\Psi^{\dagger}_R(L)\sum_iY_{i\beta}^R(L)A_i(\beta,\phi)-\Psi^{\dagger}_L(L)\sum_iY_{i\beta}^L(L)A_i(\beta,\phi)=0.\eea
Using the boundary conditions (\ref{apobc}) in the above equations, which corresponds to applying OBC, we obtain the relation between the amplitudes $A_1(\beta,\phi),A_2(\beta,\phi)$ and also the Bethe equation in the one particle sector, which are given by

\bea
\frac{A_1(\beta,\phi)}{A_2(\beta,\phi)}=\frac{\cosh\left(\frac{1}{2}(\beta+i\phi+i\pi/2)\right)}{\cosh\left(\frac{1}{2}(\beta-i\phi-i\pi/2)\right)},\eea

\bea e^{2im_0L\sinh\beta}=\frac{\cosh\left(\frac{1}{2}(\beta+i\phi+i\pi/2)\right)}{\cosh\left(\frac{1}{2}(\beta-i\phi-i\pi/2)\right)}\frac{\cosh\left(\frac{1}{2}(\beta+i\phi^{\prime}+i\pi/2)\right)}{\cosh\left(\frac{1}{2}(\beta-i\phi^{\prime}-i\pi/2)\right)}.\eea
Expressing the boundary parameters in terms of the boundary SG fields $\phi_L,\phi_R$, we have

\bea e^{2im_0L\sinh\beta}=\frac{\cosh\left(\frac{1}{2}(\beta+i\phi_L+iu)\right)}{\cosh\left(\frac{1}{2}(\beta-i\phi_L-iu)\right)}\frac{\cosh\left(\frac{1}{2}(\beta+i\phi_R+iu)\right)}{\cosh\left(\frac{1}{2}(\beta-i\phi_R-iu)\right)}.\eea
\vspace{3mm}

\subsection{Two particles sector}

Now consider the two particle case. Due to the interactions in the Hamiltonian, the ordering of the particles is important. The wavefunction in the two particle sector can be written as

\bea \ket{2}= \sum_{\alpha_1,\alpha_2=L,R}\int_{0}^{L}\int_{0}^{L} dx_1 dx_2\;\; \Psi^{\dagger}_{\alpha_1}(x_1)\Psi^{\dagger}_{\alpha_2}(x_2)\mathcal{A}\Upsilon_{\beta_1\beta_2}^{\alpha_1\alpha_2}(x_1,x_2),\label{2pwf}
\eea
where $\beta_1,\beta_2$ are the rapidities and $\mathcal{A}$ is the antisymmetrizer $\mathcal{A}\Upsilon_{\beta_1\beta_2}^{\alpha_1\alpha_2}(x_1,x_2)=\Upsilon_{\beta_1\beta_2}^{\alpha_1\alpha_2}(x_1,x_2)-\Upsilon_{\beta_1\beta_2}^{\alpha_2\alpha_1}(x_2,x_1)$.

\vspace{3mm}

Applying the Hamiltonian (\ref{apMThamiltonian}) to the two particle wavefunction (\ref{2pwf}) we obtain the following set of equations:

\bea (-i(\partial_{x_1}+\partial_{x_2})-(E_1+E_2))\mathcal{A}\Upsilon_{\beta_1\beta_2}^{RR}(x_1,x_2)-im_0\left(\mathcal{A}\Upsilon_{\beta_1\beta_2}^{RL}(x_1,x_2)+\mathcal{A}\Upsilon_{\beta_1,\beta_2}^{LR}(x_1,x_2)\right)=0,
\\
(-i(\partial_{x_1}+\partial_{x_2})-(E_1+E_2))\mathcal{A}\Upsilon_{\beta_1\beta_2}^{LL}(x_1,x_2)+im_0\left(\mathcal{A}\Upsilon_{\beta_1\beta_2}^{RL}(x_1,x_2)+\mathcal{A}\Upsilon_{\beta_1,\beta_2}^{LR}(x_1,x_2)\right)=0,\\\nonumber
(i(\partial_{x_1}-\partial_{x_2})-(E_1+E_2))\mathcal{A}\Upsilon_{\beta_1\beta_2}^{LR}(x_1,x_2)+im_0\left(-\mathcal{A}\Upsilon_{\beta_1\beta_2}^{LL}(x_1,x_2)+\mathcal{A}\Upsilon_{\beta_1,\beta_2}^{RR}(x_1,x_2)\right)\\+g\delta(x_1-x_2)\left(\mathcal{A}\Upsilon_{\beta_1\beta_2}^{LR}(x_1,x_2)-\mathcal{A}\Upsilon_{\beta_1,\beta_2}^{RL}(x_1,x_2)\right)=0,\\\nonumber
(-i(\partial_{x_1}-\partial_{x_2})-(E_1+E_2))\mathcal{A}\Upsilon_{\beta_1\beta_2}^{RL}(x_1,x_2)+im_0\left(-\mathcal{A}\Upsilon_{\beta_1\beta_2}^{LL}(x_1,x_2)+\mathcal{A}\Upsilon_{\beta_1,\beta_2}^{RR}(x_1,x_2)\right)\\+g\delta(x_1-x_2)\left(\mathcal{A}\Upsilon_{\beta_1\beta_2}^{RL}(x_1,x_2)-\mathcal{A}\Upsilon_{\beta_1,\beta_2}^{LR}(x_1,x_2)\right)=0.\label{2peq}
\eea
In order to solve the above equations, we consider the following ansatz for $\mathcal{A}\Upsilon_{\beta_1\beta_2}^{\alpha_a\alpha_2}(x_1,x_2)$:

\bea \nonumber\mathcal{A}\Upsilon_{\beta_1\beta_2}^{\alpha_a\alpha_2}(x_1,x_2)=\sum_{ij} \left(Y_{i\beta_1}^{\alpha_1}(x_1)Y_{j\beta_2}^{\alpha_2}(x_2)-Y_{i\beta_1}^{\alpha_2}(x_2)Y_{j\beta_2}^{\alpha_1}(x_1)\right)\left(A_{12}^{ij}\theta(x_1-x_2)+A_{21}^{ij}\theta(x_2-x_1)\right)\\\nonumber\theta(x_1)\theta(L-x_1)\theta(x_2)\theta(L-x_2),\\\eea
where $Y_{i\beta}^{L,R}(x)$ are given by (\ref{1pansatz}). Using the above ansatz in equations (\ref{2peq}) and applying the boundary conditions (\ref{apobc}), we obtain the Bethe equations in the two particle sector:

\bea \nonumber e^{2im_0L\sinh\beta_i}=\frac{\cosh\left(\frac{1}{2}(\beta_i+i\phi_L+iu)\right)}{\cosh\left(\frac{1}{2}(\beta_i-i\phi_L-iu)\right)}\frac{\cosh\left(\frac{1}{2}(\beta_i+i\phi_R+iu)\right)}{\cosh\left(\frac{1}{2}(\beta_i-i\phi_R-iu)\right)}\frac{\sinh\left(\frac{1}{2}(\beta_i-\beta_j+2iu)\right)}{\sinh\left(\frac{1}{2}(\beta_i-\beta_j-2iu)\right)}\frac{\sinh\left(\frac{1}{2}(\beta_i+\beta_j+2iu)\right)}{\sinh\left(\frac{1}{2}(\beta_i+\beta_j-2iu)\right)},\\\eea
where $i,j=1,2$ and $u=\pi/2-\tan^{-1}(g/2)$. The energy of the eigenstate in the two particle sector is given by $E=E_1+E_2=m_0\cosh(\beta_1)+m_0\cosh(\beta_2)$.

\subsection{Bethe equations in the N particles sector}

Similarly, one can construct $N$ particle wavefunction

\bea \nonumber\ket{N}=\sum_{\alpha_1,...\alpha_N=L,R}\int_{0}^{L}\prod_{i=1}^N dx_i \Psi^{\dagger}_{\alpha_i}(x_i)\mathcal{A}\Upsilon^{\alpha_1...\alpha_N}_{\beta_1...\beta_N}(x_1,...x_N)\ket{0},\\
\Upsilon^{\alpha_1...\alpha_N}_{\beta_1...\beta_N}(x_1,...x_N)= \sum_Q\sum_{\delta_i=1,2}\prod_{i=1}^{N} Y^{\alpha_i}_{\delta_i,\beta_i}(x_i) A^{\{\delta_i\}}[Q]\theta(Q),
\eea
where $\mathcal{A}$ represents anti-symmetrization with respect to the indices $\alpha_i,x_i$. $Q$ corresponds to different orderings of particles in the configuration space. Following the similar procedure as described above we obtain the following Bethe equations in the $N$ particle sector:

\bea \nonumber e^{2im_0L\sinh\beta_i}=\frac{\cosh\left(\frac{1}{2}(\beta_i+i\phi_L+iu)\right)}{\cosh\left(\frac{1}{2}(\beta_i-i\phi_L-iu)\right)}\frac{\cosh\left(\frac{1}{2}(\beta_i+i\phi_R+iu)\right)}{\cosh\left(\frac{1}{2}(\beta_i-i\phi_R-iu)\right)}\prod_{i\neq j,j=1}^N\frac{\sinh\left(\frac{1}{2}(\beta_i-\beta_j+2iu)\right)}{\sinh\left(\frac{1}{2}(\beta_i-\beta_j-2iu)\right)}\frac{\sinh\left(\frac{1}{2}(\beta_i+\beta_j+2iu)\right)}{\sinh\left(\frac{1}{2}(\beta_i+\beta_j-2iu)\right)}.\\\label{BEMT}\eea
An eigenstate of the Hamiltonian corresponds to a unique set of the Bethe roots $\{\beta_j\}$, which are solutions to the Bethe equations (\ref{BEMT}). The energy of the eigenstate is given by
\bea E=\sum_{j=1}^N m_0\cosh(\beta_j).\label{eneq}
\eea
Recall that the above Bethe equations correspond to $m_0>0$ and OBC (\ref{apobc}). Had we instead considered the case of $m_0<0$ with OBC (\ref{apobc}), or equivalently the case of $m_0>0$ with $\widehat{\text{OBC}}$ (\ref{aptobc}), we would have obtained the following set of Bethe equations:

\bea \nonumber e^{2im_0L\sinh\beta_i}=\frac{\sinh\left(\frac{1}{2}(\beta_i+i\phi_L+iu)\right)}{\sinh\left(\frac{1}{2}(\beta_i-i\phi_L-iu)\right)}\frac{\sinh\left(\frac{1}{2}(\beta_i+i\phi_R+iu)\right)}{\sinh\left(\frac{1}{2}(\beta_i-i\phi_R-iu)\right)}\prod_{i\neq j,j=1}^N\frac{\sinh\left(\frac{1}{2}(\beta_i-\beta_j+2iu)\right)}{\sinh\left(\frac{1}{2}(\beta_i-\beta_j-2iu)\right)}\frac{\sinh\left(\frac{1}{2}(\beta_i+\beta_j+2iu)\right)}{\sinh\left(\frac{1}{2}(\beta_i+\beta_j-2iu)\right)}.\\\label{BEMTtr}\eea
The energy of the eigenstates is the same as before (\ref{eneq}). As we shall demonstrate, the solutions to the above Bethe equations (\ref{BEMT}), (\ref{BEMTtr}) correspond to the states with negative charge. To obtain the states with positive charge one needs to consider charge conjugated fermions. Under the charge conjugation 

\be\mathcal{C}\Psi^{\dagger}_{L,R}=\Psi^{'\dagger}_{L,R}(x)=\Psi_{L,R}(x),\label{apcc}\ee 
the Hamiltonian remains invariant. However the boundary conditions on the fermion 
fields $\Psi^{'}_{L,R}$ are not the complex conjugates of (\ref{apobc}),(\ref{aptobc}). Rather, we find that  charge conjugation affects only the physical boundary fields $\phi_{L,R}\rightarrow -\phi_{L,R}$ but keeps the anomalous term intact, i.e the OBC (\ref{apobc}) and (\ref{aptobc})now takes the form:

\bea
\Psi^{'}_L(0)=e^{i\phi}\Psi^{'}_R(0), \hspace{4mm} \Psi^{'}_L(L)=-e^{-i\phi^{\prime}}\Psi^{'}_R(L),\\
\phi=-\phi_L+u-\pi/2, \;\; \phi^{\prime}=-\phi_R+u-\pi/2.\label{ccbc}
\eea
and 
\bea
\Psi^{'}_L(0)=e^{i\phi}\Psi^{'}_R(0), \hspace{4mm} \Psi^{'}_L(L)=-e^{-i\phi^{\prime}}\Psi^{'}_R(L),\\
\phi=-\phi_L+u+\pi/2, \;\; \phi^{\prime}=-\phi_R+u+\pi/2\label{cctbc}
\eea
respectively. As mentioned in the main text, we find that the system is invariant under the combined application of charge conjugation and changing the sign of the boundary fields. Just as in the case of the original fermions $\Psi_{L,R}(x)$ considered above, one needs to construct the Bethe wavefunction for the charge conjugated fermions $\Psi^{'}_{L,R}(x)$ with their corresponding boundary conditions given in (\ref{ccbc}), and thus obtain the associated Bethe equations. We skip the details and directly present the Bethe equations as the procedure is exactly same as the one presented above. We have for the case of $m_0>0$ and OBC (\ref{ccbc}):

\bea \nonumber e^{2im_0L\sinh\beta_i}=\frac{\cosh\left(\frac{1}{2}(\beta_i-i\phi_L+iu)\right)}{\cosh\left(\frac{1}{2}(\beta_i+i\phi_L-iu)\right)}\frac{\cosh\left(\frac{1}{2}(\beta_i-i\phi_R+iu)\right)}{\cosh\left(\frac{1}{2}(\beta_i+i\phi_R-iu)\right)}\prod_{i\neq j,j=1}^N\frac{\sinh\left(\frac{1}{2}(\beta_i-\beta_j+2iu)\right)}{\sinh\left(\frac{1}{2}(\beta_i-\beta_j-2iu)\right)}\frac{\sinh\left(\frac{1}{2}(\beta_i+\beta_j+2iu)\right)}{\sinh\left(\frac{1}{2}(\beta_i+\beta_j-2iu)\right)}.\\\label{BEMTCC}\eea
For the case of $m_0<0$ with OBC (\ref{ccbc}), or equivalently for the case of $m_0>0$ with $\widehat{\text{OBC}}$ (\ref{cctbc}), we obtain the following set of Bethe equations:

\bea \nonumber e^{2im_0L\sinh\beta_i}=\frac{\sinh\left(\frac{1}{2}(\beta_i-i\phi_L+iu)\right)}{\sinh\left(\frac{1}{2}(\beta_i+i\phi_L-iu)\right)}\frac{\sinh\left(\frac{1}{2}(\beta_i-i\phi_R+iu)\right)}{\sinh\left(\frac{1}{2}(\beta_i+i\phi_R-iu)\right)}\prod_{i\neq j,j=1}^N\frac{\sinh\left(\frac{1}{2}(\beta_i-\beta_j+2iu)\right)}{\sinh\left(\frac{1}{2}(\beta_i-\beta_j-2iu)\right)}\frac{\sinh\left(\frac{1}{2}(\beta_i+\beta_j+2iu)\right)}{\sinh\left(\frac{1}{2}(\beta_i+\beta_j-2iu)\right)}.\\\label{BEMTcctr}\eea
The energy of the eigenstates is the same as before (\ref{eneq}). The charge of a state corresponding to the charge conjugated fermions can be expressed in terms of the original fermions as
\be \label{cchargedef}\sum_{i=L,R}\int \Psi_{i}^{'\dagger}(x)\Psi_{i}^{'}(x)=\text{Const}-\sum_{i=L,R}\int \Psi_{i}^{\dagger}(x)\Psi_{i}^{}(x).\ee
As we can see, up to an unimportant constant the charge of the charge conjugated fermions $\Psi^{'}_{L,R}(x)$ is exactly opposite to that of the original fermions  $\Psi_{L,R}(x)$.

\section{Solutions to the Bethe equations}
\label{apB}
As mentioned in the main text, in this work we focus on the repulsive regime, which corresponds to $u<\pi/2$. Each eigenstate of the Hamiltonian corresponds to a unique set of roots $\{\beta_j\}$ called Bethe roots, which satisfy the Bethe equations (\ref{BEMT},\ref{BEMTtr},\ref{BEMTCC},\ref{BEMTcctr}). Due to the presence of open boundary conditions, the Bethe equations are reflective symmetric: $\beta_i\leftrightarrow -\beta_i$, due to which the solutions to  (\ref{BEMT},\ref{BEMTtr},\ref{BEMTCC},\ref{BEMTcctr}) come in pairs $(\beta_i,-\beta_i)$ resulting in doubling of solutions. As we shall discuss below, one can avoid the double counting of the Bethe roots by introducing a counter term. In the thermodynamic limit $N,L\rightarrow\infty$, Bethe roots form a dense set and the Bethe equations can be expressed as integral equations, which can be solved using Fourier transform. As we shall see, in this limit one also needs to introduce a cutoff $\Lambda=\pi N/L\gg m_0$ on $\beta_j$. From the relation between the Bethe roots $\beta_j$ and the energy (\ref{eneq}), we can infer that the roots lying on the line $\beta_j=\alpha_j+i\pi$ have lower energy. Although this statement is in general true, one needs to be careful in accounting for the boundary bound states which are described by purely imaginary solutions to the Bethe equations called boundary strings. In this work we present our solution, given the OBC (\ref{apobc}), in two regimes of the boundary fields $\phi_{L,R} \in (-\pi, \pi]$
\bea
{\rm I} &:& -u< \phi_{L,R} <u \\
{\rm II} &:&-\pi<\phi_{L,R}<-(\pi-u),\; \pi-u<\phi_{L,R}<\pi.
\label{domains}
\eea
In the range $(-\pi, \pi]$, these two regions are mapped onto each other by the duality transformation (\ref{apduality}):  $\phi_{L,R}\rightarrow \phi_{L,R} + \pi$. When $m_0<0$ and the boundary fields lie in region I, which is described by the Bethe equations (\ref{BEMTtr},\ref{BEMTcctr}), the ground state is non degenerated and there exist no boundary bound states. We refer to this phase as the trivial phase. In contrast, when boundary fields lie in the region ${\rm II}$, which is described by the Bethe equations (\ref{BEMT},\ref{BEMTCC}), the system hosts exponentially localized boundary bound states at each edge. These are related to the existence of what we call  `close or short boundary string' (\cite{Parmeshthesis},\cite{YSRKondo}) solutions of the Bethe equations. As a result, there exists, on top of the ground state, three other states within the one particle (soliton) gap. We call  this phase ``mid gap phase". The situation for the $m_0>0$ case is reversed and can be obtained with use of the duality transformation, where the region ${\rm I}$ corresponds to mid gap phase whereas region ${\rm II}$ corresponds to trivial phase in the sense described above. When the boundary fields $\phi_{L,R}$ lie outside the regions ${\rm I}$ and ${\rm II}$, the Bethe equations exhibit new type of Bethe roots that are called  `wide boundary strings'  (\cite{Parmeshthesis},\cite{YSRKondo},\cite{XXZterras})  and the associated boundary physics is very different from that of the mid gap and trivial phases described above. The analysis in this regime is rather involved and will be presented in a further publication. Below we shall provide the solutions to the Bethe equations in the trivial phases and mid gap phases.

\subsection{Trivial Phase}
As mentioned above, the Bethe equations in the trivial phase are given by (\ref{BEMTcctr}),(\ref{BEMTtr}). Since we are working with $m_0<0$, as mentioned before, the trivial phase corresponds to the case where the boundary fields lie in the region I, where the boundary fields take values in the range $-u<\phi_{L,R}<u$. Let us first consider the case where the boundary fields $\phi_{L,R}>0$. As mentioned before, from the relation between the Bethe roots $\beta_j$ and the energy (\ref{eneq}), we can infer that the roots lying on the line $\beta_j=\alpha_j+i\pi$ have lower energy. Since we are interested in the low energy spectrum, we consider the state which has all the roots lying on this line.  Consider the Bethe equations corresponding to the charge conjugated fermions (\ref{BEMTcctr}). By making the transformation $\beta_j\rightarrow\alpha_j+i\pi$ and applying logarithm, we obtain the logarithmic form of the Bethe equations

\bea \nonumber-2im_0L\sinh\alpha_i=\ln\left(\frac{\cosh\left(\frac{1}{2}(\alpha_i-i\phi_L+iu)\right)}{\cosh\left(\frac{1}{2}(\alpha_i+i\phi_L-iu)\right)}\right)+\ln\left(\frac{\cosh\left(\frac{1}{2}(\alpha_i-i\phi_R+iu)\right)}{\sinh\left(\frac{1}{2}(\alpha_i+i\phi_R-iu)\right)}\right)\\-2i\pi n_i+\ln\left(\frac{\sinh\left(\frac{1}{2}(\alpha_i-iu)\right)}{\sinh\left(\frac{1}{2}(\alpha_i+iu)\right)}\right)-\sum_{\sigma=\pm}\sum_j\ln\left(\frac{\sinh\left(\frac{1}{2}(\alpha_i-\sigma\alpha_j-2iu)\right)}{\sinh\left(\frac{1}{2}(\alpha_i-\sigma\alpha_j+2iu)\right)}\right).\label{logBEcctr}\eea
Here $n_i$ are integers corresponding to the logarithmic branch. These integers are in one to one correspondence with the Bethe roots $\alpha_i$. Note that the selection rule $i\neq j$ which is necessary for non vanishing of the wave function has been lifted by adding a counter term that cancels the term arising from $\alpha_i=\alpha_j$. $\alpha_i=0$ is a trivial solution to (\ref{BEMTtr}) and this leads to a vanishing wave function \cite{skorik}. Hence, this solution needs to be removed, and this can be achieved by omitting the integer $n_i$ corresponding to $\alpha_i=0$. As we shall see, this leads to a delta function centered at $\alpha=0$ in the density distribution of roots to be defined below.  
\vspace{3mm}

Using the notation $\varphi(x,y)=i\ln\left(\frac{(\sinh(x+iy)/2)}{(\sinh(x-iy)/2)}\right)$, the above equation can be written as

\bea 2m_0L\sinh\alpha_i=\varphi(\alpha_i,-\phi_R+u+i\pi)+\varphi(\alpha_i,-\phi_L+u+i\pi)-\sum_{\sigma=\pm}\sum_{j}\varphi(\alpha_i-i+\sigma\alpha_j,2u)-\varphi(2\alpha_i,2u)+2\pi n_i.\label{belgtr}\eea
Subtracting the equation (\ref{belgtr}) for the root $\alpha_{i+1}$ from that of the root $\alpha_i$, we have

\bea \nonumber & 2m_0L\left(\sinh\alpha_{i+1}-\sinh\alpha_i\right)=\varphi\left(\alpha_{i+1},-\phi_R+u+i\pi\right)-\varphi\left(\alpha_{i},-\phi_R+u+i\pi\right) + \varphi(\alpha_{i+1},-\phi_L+u+i\pi)\\\nonumber&-\varphi(\alpha_i,-\phi_L+u+i\pi)-(\varphi(2\alpha_{i+1},2u)-\varphi(2\alpha_i,2u))-\sum_{\sigma=\pm}\sum_j (\varphi(\alpha_{i+1}+\sigma\alpha_j,2u)-\varphi(\alpha_i+\sigma\alpha_j,2u))+2\pi(n_{i+1}-n_i).\\\label{beeqdifftr}
\eea
In this state the integers $n_i$ are consecutively filled without any gap except for the integer corresponding to $\alpha_i=0$ as mentioned above. In the thermodynamic limit as mentioned above, Bethe roots form a dense set. In this limit one can define the density distribution of roots as

\be\label{defden} \rho(\alpha)=\frac{1}{2L}\frac{1}{(\alpha_{i+1}-\alpha_i)},
\ee
where the factor of $2$ is to account for the doubling of the solutions mentioned above. Using this in the equation (\ref{beeqdifftr}), we have

\bea \nonumber & 2m_0\cosh\alpha-\frac{1}{L}\varphi^{\prime}(\alpha,-\phi_R+u+i\pi)-\frac{1}{L}\varphi^{\prime}(\alpha,-\phi_L+u+i\pi)-4\pi\rho_{\ket{0}}(\alpha)-\frac{1}{L}\delta(\alpha)+\frac{1}{L}\varphi^{\prime}(\alpha,u)+\frac{1}{L}\varphi^{\prime}(\alpha,u+i\pi)\\&=\sum_{\sigma}\sum_j\frac{1}{L}\varphi^{\prime}(\alpha+\sigma\gamma_j,2u)=\sum_{\sigma}\int\varphi^{\prime}(\alpha+\sigma\gamma,2u)\rho_{\ket{0}}(\gamma)d\gamma,
\label{gsdentr}\eea
where \be\varphi^{\prime}(x,y)=\partial_x\varphi(x,y)=\frac{\sin y}{\cosh x-\cos y},\;\;\;\varphi^{\prime}(x,y+i\pi)=\partial_x\varphi(x,y+i\pi)=-\frac{\sin y}{\cosh x+\cos y}.\ee
Note that we labelled the density distribution by $\rho_{\ket{0}}(\alpha)$. The reason for this labelling will be discussed shortly. The above equation (\ref{gsdentr}) can be solved by applying Fourier transform. The Fourier transforms of all the terms in the above equation are well defined except for the the $\cosh(\alpha)$ term. In order to tackle this, we need to apply a cutoff on the `rapidity' values $\alpha$ when taking the Fourier transform of this term \cite{Thacker}. We obtain

\bea\tilde{\rho}_{\ket{0}}(\omega)=\frac{2m_0\tilde{c}(\omega)-\frac{1}{L}\sum_{a=R,L}\tilde{\varphi}^{\prime}(\omega,-\phi_a+u+i\pi)+\frac{1}{L}\tilde{\varphi}^{\prime}(\omega,u)+\frac{1}{L}\tilde{\varphi}^{\prime}(\omega,u+i\pi)-1}{2(2\pi+\tilde{\varphi}^{\prime},(\omega,2u))},
\eea
where \bea\tilde{c}(\omega)=\int_{-\Lambda}^{\Lambda}d\alpha\;e^{i\alpha\omega}\cosh(\alpha)=\frac{e^{\Lambda}}{2}\left(\frac{e^{i\Lambda \omega}}{1+i\omega}+ \frac{e^{-i\Lambda \omega}}{1-i\omega}\right),\\\tilde{\varphi}^{\prime}(\omega,y)=\int_{-\infty}^{\infty}dx\;e^{ix\omega}\varphi^{\prime}(x,y)=2\pi\frac{\sinh((\pi-y)\omega)}{\sinh(\pi\omega)},\\\tilde{\varphi}^{\prime}(\omega ,y+i\pi)=\int_{-\infty}^{\infty}dx\;e^{ix\omega}\varphi^{\prime}(x,y+i\pi)=-2\pi\frac{\sinh(y\omega)}{\sinh(\pi\omega)}. \; (y<\pi). \label{FTforms}\eea
Explicitly, we have

\bea \label{dentr}\tilde{\rho}_{\ket{0}}(\omega)=\tilde{\rho}_{Bulk}(\omega)+\tilde{\rho}_{Boundary}(\omega)+\tilde{\rho}^{tr}_{Boundary fields}(\omega),\eea
where
\bea\label{denbulk}\tilde{\rho}_{Bulk}(\omega)=\frac{m_0}{4\pi}\;\frac{\sinh(\pi\omega)}{\sinh(\pi\omega)+\sinh((\pi-2u)\omega)}\;e^{\Lambda}\left(\frac{e^{i\Lambda \omega}}{1+i\omega}+ \frac{e^{-i\Lambda \omega}}{1-i\omega}\right),\eea

\bea \label{denb}\tilde{\rho}_{Boundary}(\omega)=\frac{1}{2L}\frac{\sinh((\pi-u)\omega)-\sinh(u\omega)-\sinh(\pi\omega)}{\sinh(\pi\omega)+\sinh((\pi-2u)\omega)},\eea

\bea \label{denbtr} \tilde{\rho}^{tr}_{Boundary fields}(\omega)=\frac{1}{2L}\sum_{a=L,R}\frac{\sinh((u-\phi_a))\omega)}{\sinh(\pi\omega)+\sinh((\pi-2u)\omega)}.\eea
The charge of the state corresponding to the charge conjugated fermions obtained above by filling the roots continuously on the $i\pi$ line, which is described by the density distribution $\rho_{\ket{0}}(\alpha)$ is 

\be \label{tchargetr} -L\int_{-\Lambda}^{\Lambda} d\alpha\;\rho_{\ket{0}}(\alpha)= -L\tilde{\rho}_{\ket{0}}(0)=-\left(\frac{m_0 L e^{\Lambda}}{2\pi}\;\frac{\pi}{2(\pi-u)}-\frac{\phi_L+\phi_R}{4(\pi-u)}\right).
\ee
Note that the overall (-) sign arises since we are expressing the charge of the charge conjugated fermions in terms of the original fermions (\ref{cchargedef}). The first term which is divergent in the limit $\Lambda\rightarrow\infty$ is the charge associated with the bulk. The second and the third terms are due to the charge associated with the boundaries. This can be expressed as

\be\label{cformtr} -L\tilde{\rho}_{\ket{0}}(0)= -q\left(\frac{m_0 L e^{\Lambda}}{2\pi}-\frac{\phi_L+\phi_R}{2\pi}\right), \;\; \;q=\frac{\pi}{2(\pi-u)},\ee
where $q$ is equal to the renormalized charge of the soliton \cite{Thacker}. At this point we pause to stress  the importance of the anomalous terms in the boundary conditions (\ref{apobc})(\ref{aptobc}). Indeed,  without the anomalous term in the boundary condition, one would not have been able to  factor out the renormalized soliton charge $q$ in the expression of the total charge (\ref{cform-1}). The later fact  is reminiscent of   field renormalization in the MT model, i.e: $\psi= \sqrt{q}\; \psi_{{\rm phys}} $, which relates the bare fermion field $\psi$ to  the physical  fermion field $\psi_{{\rm phys}}$ associated with physical excitations.

\vspace{3mm}

\paragraph{Normal ordering.} Since the total charge is divergent   normal ordering is needed.  We define the normal ordered  charge $\mathcal{N}$ (in units of the renormalized charge of the physical excitations $q$)  of the state as
\bea \mathcal{N}= \frac{1}{q}\left\{\tilde{\rho}_{\ket{0}}(0)-\tilde{\rho}_{Bulk}(0)\right\}. 
\eea
With this definition,  the normal ordered charge of the state obtained above is given by:
\be\mathcal{N}=\frac{\phi_L+\phi_R}{2\pi}.\label{chargetr}\ee
The normal ordering prescription defined above is not unique. Indeed, as it appears, the boundary field dependence in (\ref{chargetr}) also occurs when calculating the charge of all states (see Eqs. (\ref{charge1}), (\ref{charge-1}), (\ref{charge-1A2}), (\ref{charge-1A22}), (\ref{charge+1A2}), (\ref{charge+1A22}), (\ref{charge0La2}), (\ref{charge0LA22}), (\ref{charge0LRa1}), (\ref{charge0RA22}) below). Hence one may also have consistently defined the normal ordered charge of all states as 
\be
\mathcal{N}\rightarrow  \mathcal{N} -\frac{\phi_L+\phi_R}{2\pi}\label{Nprime}.\ee
With the later definition the charges of all states do not depend on the boundary fields, and hence $\mathcal{N}$ is always an integer which can be used to label all states. This eventually allows us  to define a fermionic parity operator as 
\be
{\cal P}= (-1)^{\mathcal{N}}.
\label{fermionparity}
\ee
By this definition, we see that the charge of the state described by the density distribution $\rho_{\ket{0}}(\alpha)$ (\ref{dentr}) is
\be\mathcal{N}=0.\label{chargetrstate}
\ee
This allows us to label this state as $\ket{0}$. This is the unique ground state exhibited by the system. Hence we find that the above described solution to the Bethe equations corresponding to the charge conjugated fermions gives a state which we labelled $\ket{0}$. At this point one may ask what would have been the result if we had instead considered the Bethe equations of the original fermions (\ref{BEMTtr}). More precisely, can we obtain the state $\ket{0}$ using the Bethe equations of the original fermions (\ref{BEMTtr}). As we shall demonstrate this is not possible as the charge of the resultant state would exceed the maximum allowed charge for any state, which is

\be \label{MaxC}\text{Maximum charge}=\pm\tilde{\rho}_{Bulk}(0)=\pm\frac{m_0 L e^{\Lambda}}{2\pi}. \ee
Here the $(\pm)$ signs correspond to the states corresponding to the original and charge conjugated fermions respectively. The maximum charge (\ref{MaxC}) is equal to the charge associated with the state $\ket{0}$ obtained above when the boundary fields $\phi_{L,R}=0$, and it is also equal to the charge of the ground state when periodic boundary conditions are applied \cite{Thacker}, where all the Bethe roots on the $i\pi$ line are continuously filled without any holes.  There exists an analogous quantity for the maximum charge (\ref{MaxC}) in spin chains and field theories with internal symmetry. For example, consider the case of the spin 1/2 Heisenberg chain with $N$ sites. The reference state with all spins pointing in the downward (upward) direction correspond to the vaccuum corresponding to the original fermions (charge conjugated fermions) of the Massive Thirring model. The state with maximum charge (\ref{MaxC}) corresponds to a state with $N/2$ spins flipped in the Heisenberg chain, i.e., a state with $S^z=0$. In the spin chain, one cannot flip more spins or add more Bethe roots to the state with $S^z=0$ as one cannot cross the `equator', which in the Massive Thirring model corresponds to exceeding the maximum charge (\ref{MaxC}).

\vspace{3mm}

Here we shall demonstrate that by considering the Bethe equations corresponding to the original fermions (\ref{BEMTtr}) in the regime where $\phi_{L,R}\in (0,u)$, and filling all the roots on the $i\pi$ line as in the previous case, one obtains a state which exceeds the maximum charge (\ref{MaxC}), and hence is not a valid state.  By making the transformation $\beta_j\rightarrow\alpha_j+i\pi$ in (\ref{BEMTtr}) and applying logarithm, we obtain the logarithmic form of the Bethe equations

\bea \nonumber-2im_0L\sinh\alpha_i=\ln\left(\frac{\cosh\left(\frac{1}{2}(\alpha_i+i\phi_L+iu)\right)}{\cosh\left(\frac{1}{2}(\alpha_i-i\phi_L-iu)\right)}\right)+\ln\left(\frac{\cosh\left(\frac{1}{2}(\alpha_i+i\phi_R+iu)\right)}{\sinh\left(\frac{1}{2}(\alpha_i-i\phi_R-iu)\right)}\right)\\-2i\pi n_i+\ln\left(\frac{\sinh\left(\frac{1}{2}(\alpha_i-iu)\right)}{\sinh\left(\frac{1}{2}(\alpha_i+iu)\right)}\right)-\sum_{\sigma=\pm}\sum_j\ln\left(\frac{\sinh\left(\frac{1}{2}(\alpha_i-\sigma\alpha_j-2iu)\right)}{\sinh\left(\frac{1}{2}(\alpha_i-\sigma\alpha_j+2iu)\right)}\right).\label{logBEtr}\eea
These equations can be solved by following the same procedure described above. We denote the density distribution of the resulting state with $\rho^{\text{incorrect}}_{\ket{0}}$. We obtain

\bea \tilde{\rho}^{\text{incorrect}}_{\ket{0}}(\omega)=\tilde{\rho}_{Bulk}(\omega)+\tilde{\rho}_{Boundary}(\omega)+\tilde{\rho}^{tr}_{Boundary fields}(\omega),\eea
where $\tilde{\rho}_{Bulk}(\omega)$ and $\tilde{\rho}_{Boundary}(\omega)$ are the same as before (\ref{denbulk}), (\ref{denb}) respectively, and $\tilde{\rho}^{tr}_{Boundary fields}(\omega)$ takes the form of (\ref{denbtr}) with the change $\phi_{L,R}\rightarrow -\phi_{L,R}$. Explicitly we have

\bea \label{denbtr} \tilde{\rho}^{tr'}_{Boundary fields}(\omega)=\frac{1}{2L}\sum_{a=L,R}\frac{\sinh((u+\phi_a))\omega)}{\sinh(\pi\omega)+\sinh((\pi-2u)\omega)}.\eea
The charge of this state is 

\be \label{tchargetrinc} L\int_{-\Lambda}^{\Lambda} d\alpha\;\rho^{\text{incorrect}}_{\ket{0}}(\alpha)= L\tilde{\rho}^{\text{incorrect}}_{\ket{0}}(0)=\left(\frac{m_0 L e^{\Lambda}}{2\pi}\;\frac{\pi}{2(\pi-u)}+\frac{\phi_L+\phi_R}{4(\pi-u)}\right), \;\; \phi_{L,R}\in (0,u).
\ee
As we can see, for finite $\phi_{L,R}\neq 0$, the total charge of the state described by the density distribution $\rho^{\text{incorrect}}_{\ket{0}}(\alpha)$ exceeds the maximum charge (\ref{MaxC}). Hence this is not an allowed state as mentioned above. Hence we see that for boundary fields $\phi_{L,R}$ taking values in the range $(0,u)$, a valid solution is obtained by filling all the roots on the $i\pi$ line only when the charge conjugated Bethe equations are considered.

\vspace{3mm}

Now consider the case where the boundary fields take values in the range $\phi_{L,R}\in (-u,0)$. It is easier to work with absolute values of the boundary fields, and hence let us make the following transformation $\phi_{L,R}\rightarrow -\phi^{'}_{L,R}$, where $\phi^{'}_{L,R}\equiv |\phi_{L,R}|$. The Bethe equations corresponding to the original fermions (\ref{BEMTtr}) take the following form

\bea \nonumber e^{2im_0L\sinh\beta_i}=\frac{\sinh\left(\frac{1}{2}(\beta_i-i\phi^{'}_L+iu)\right)}{\sinh\left(\frac{1}{2}(\beta_i+i\phi_L^{'}-iu)\right)}\frac{\sinh\left(\frac{1}{2}(\beta_i-i\phi^{'}_R+iu)\right)}{\sinh\left(\frac{1}{2}(\beta_i+i\phi^{'}_R-iu)\right)}\prod_{i\neq j,j=1}^N\frac{\sinh\left(\frac{1}{2}(\beta_i-\beta_j+2iu)\right)}{\sinh\left(\frac{1}{2}(\beta_i-\beta_j-2iu)\right)}\frac{\sinh\left(\frac{1}{2}(\beta_i+\beta_j+2iu)\right)}{\sinh\left(\frac{1}{2}(\beta_i+\beta_j-2iu)\right)}.\\\eea
Considering the Bethe roots on the $i\pi$ line, and applying the logarithm to the above equations we have

\bea \nonumber-2im_0L\sinh\alpha_i=\ln\left(\frac{\cosh\left(\frac{1}{2}(\alpha_i-i\phi^{'}_L+iu)\right)}{\cosh\left(\frac{1}{2}(\alpha_i+i\phi^{'}_L-iu)\right)}\right)+\ln\left(\frac{\cosh\left(\frac{1}{2}(\alpha_i-i\phi^{'}_R+iu)\right)}{\sinh\left(\frac{1}{2}(\alpha_i+i\phi^{'}_R-iu)\right)}\right)\\-2i\pi n_i+\ln\left(\frac{\sinh\left(\frac{1}{2}(\alpha_i-iu)\right)}{\sinh\left(\frac{1}{2}(\alpha_i+iu)\right)}\right)-\sum_{\sigma=\pm}\sum_j\ln\left(\frac{\sinh\left(\frac{1}{2}(\alpha_i-\sigma\alpha_j-2iu)\right)}{\sinh\left(\frac{1}{2}(\alpha_i-\sigma\alpha_j+2iu)\right)}\right).\label{logBEtr+}\eea
These equations are exactly same as the Bethe equations corresponding to the charge conjugated fermions in the regime $\phi_{L,R}\in(0,u)$ upto the change $\phi_{L,R}\leftrightarrow \phi_{L,R}^{'}$. They can be solved following the same procedure as above. Denoting the density distribution of the resulting state by $\rho_{\ket{0}}(\alpha)$, we have 

\bea \label{dentr+}\tilde{\rho}_{\ket{0}}(\omega)=\tilde{\rho}_{Bulk}(\omega)+\tilde{\rho}_{Boundary}(\omega)+\tilde{\rho}^{tr}_{Boundary fields}(\omega),\eea
where $\tilde{\rho}_{Bulk}(\omega)$ and $\tilde{\rho}_{Boundary}$ are given by (\ref{denbulk}) and (\ref{denb}) respectively, and 

\bea \label{denbtr} \tilde{\rho}^{tr}_{Boundary fields}(\omega)=\frac{1}{2L}\sum_{a=L,R}\frac{\sinh((u-\phi^{'}_a))\omega)}{\sinh(\pi\omega)+\sinh((\pi-2u)\omega)}.\eea
The charge of this state is 

\be \label{tchargetr+} L\int_{-\Lambda}^{\Lambda} d\alpha\;\rho_{\ket{0}}(\alpha)= L\tilde{\rho}_{\ket{0}}(0)=\left(\frac{m_0 L e^{\Lambda}}{2\pi}\;\frac{\pi}{2(\pi-u)}-\frac{\phi^{'}_L+\phi^{'}_R}{4(\pi-u)}\right), \;\; \phi^{'}_{L,R}\in (0,u).
\ee

Recall that $\phi^{'}_{L,R}\equiv |\phi_{L,R}|$.  We see that the charge of this state is less than the maximum allowed charge (\ref{MaxC}), and therefore we have obtained a valid solution by considering the Bethe equations of the original fermions and filling all the roots on the $i\pi$ line. Just as before we define the normal ordered charge $\mathcal{N}$ (in units of the renormalized charge of the physical excitations $q$)  of the state corresponding to the original fermions as
\bea \mathcal{N}= \frac{1}{q}\left\{\tilde{\rho}_{\ket{0}}(0)-\tilde{\rho}_{Bulk}(0)\right\}. 
\eea
With this definition,  the normal ordered charge of the state obtained above is given by:
\be\mathcal{N}=-\frac{\phi^{'}_L+\phi^{'}_R}{2\pi}.\label{chargetr+}\ee
Reverting back to the original boundary fields, we have

\be\mathcal{N}=\frac{\phi_L+\phi_R}{2\pi}.\label{chargetr+or}\ee
Just as before (\ref{Nprime}), one can absorb the boundary fields into the normal ordered charge 
\be
\mathcal{N}\rightarrow  \mathcal{N} -\frac{\phi_L+\phi_R}{2\pi}\label{Nprime+},\ee
such that the normal ordered charge of the state is
\be\mathcal{N}=0.
\ee

Had we instead considered the Bethe equations corresponding to the charge conjugated fermions with all roots lying on the $i\pi$ line, we would have obtained a state whose charge would have exceeded the maximum allowed charge (\ref{MaxC}) just as in (\ref{tchargetrinc}). Therefore, we end up with the following picture:  In the trivial phase when boundary fields $\phi_{L,R}\in (0,u)$, the ground state is obtained by considering the Bethe equations corresponding to the charge conjugated fermions whereas for $\phi_{L,R}\in (-u,0)$, one should instead consider the Bethe equations corresponding to the original fermions. In both cases the all the Bethe roots lie on the $i\pi$ line. The normal ordered charge of the state is $\mathcal{N}=0$.  Now consider the case where one boundary field lies in the range $(0,u)$, whereas the other boundary field lies in the range $(-u,0)$. Looking at the form of the total charge (\ref{tchargetr},\ref{tchargetr+}), we can infer that when the sum of the boundary fields is positive, one needs to work with the Bethe equations corresponding to the charge conjugated fermions, whereas when it is negative one needs to work with the Bethe equations corresponding to the original fermions. Just as before, the ground state is obtained by considering all the roots on the $i\pi$ line and it is unique with normal ordered charge $\mathcal{N}=0$.
Hence, in the trivial phase when $m_0<0$ for all values of the boundary fields in the range $(-u,u)$, the ground state is unique with total normal ordered charge $\mathcal{N}=0$ and hence it is labelled as $\ket{0}$.

\subsubsection{Solitons}  The simplest excitations in the bulk constitute of solitons. A soliton can be created on top of a state by removing a root from the density distribution corresponding to that particular state. This is also referred to as adding a hole. Let us consider the case of $\phi_{L,R}\in (0,u)$ and remove a root from the density distribution corresponding to the state $\ket{0}$ described above. Removing a root is equivalent to omitting an integer $n_i$ in (\ref{logBEcctr}) corresponding to that particular root. As mentioned before, an omitted integer gives rise to a delta function in the density distribution. Consider a state with one hole at rapidity $\alpha=\theta$ (or equivalently removing the root $\theta$). Denoting the density distribution of the resulting state by $\rho_{\theta}(\alpha)$, we have

\bea \nonumber & 2m_0\cosh\alpha-\frac{1}{L}\varphi^{\prime}(\alpha,-\phi_R+u+i\pi)-\frac{1}{L}\varphi^{\prime}(\alpha,-\phi_L+u+i\pi)-4\pi\rho_{\theta}(\alpha)-\frac{1}{L}\delta(\alpha)+\frac{1}{L}\varphi^{\prime}(\alpha,u)+\frac{1}{L}\varphi^{\prime}(\alpha,u+i\pi)\\&-\frac{1}{L}\sum_{\sigma=\pm}\delta(\alpha+\sigma\theta)=\sum_{\sigma}\int\varphi^{\prime}(\alpha+\sigma\gamma,2u)\rho_{\theta}(\gamma)d\gamma.
\label{gsdenhole}\eea

This equation can be solved by following the same procedure described above. We obtain

\be \tilde{\rho}_{\theta}(\omega)=\tilde{\rho}_{\ket{0}}(\omega)+\Delta\tilde{\rho}_{\theta}, \;\;\; \Delta\tilde{\rho}_{\theta}(\omega)=-\frac{1}{2L}\;\frac{\sinh(\pi\omega)\cos(\theta\omega)}{\sinh((\pi-u)\omega)\cosh(u\omega)}\label{denhole},
\ee
where $\tilde{\rho}_{\ket{0}}(\omega)$ is given by (\ref{dentr}). We see that by adding a soliton (hole) with rapidity $\alpha=\theta$ to the state $\ket{0}$, the associated density distribution $\rho_{\ket{0}}(\alpha)$ undergoes a change $\Delta\rho_{\theta}(\alpha)$, resulting in a new density distribution $\rho_{\theta}(\alpha)$. The charge associated with the state described by this distribution is 
\be -L\tilde{\rho}_{\theta}(0)=-L\tilde{\rho}_{\ket{0}}(0)-L\Delta\tilde{\rho}_{\theta}(0)=-L\tilde{\rho}_{\ket{0}}(0)+\frac{\pi}{2(\pi-u)}.\label{chargeholetr+form}
\ee
Using the expression for the charge associated with the state $\ket{0}$ (\ref{dentr}) in the equation (\ref{chargeholetr+form}), we find that the charge associated with the state containing the soliton is

 \be -L\tilde{\rho}_{\theta}(0)= -q\left(\frac{m_0 L e^{\Lambda}}{2\pi}-1-\frac{\phi_L+\phi_R}{2\pi}\right), \;\; \;q=\frac{\pi}{2(\pi-u)}.\label{chargeholestate}\ee
Defining the normal ordered charge as before, we find that the normal ordered charge of the state containing a soliton on top of the state $\ket{0}$ is 

\be \mathcal{N}=1+\frac{\phi_L+\phi_R}{2\pi}.\ee
Just as before (\ref{Nprime}),  we can absorb the boundary fields by redefining the normal ordering. Hence, the charge of the state obtained above is
\be\mathcal{N}=1.
\ee
One defines the charge of a soliton as the change in the charge of a state due to the addition of a hole. Comparing the charges of the state with and without the hole, i.e., (\ref{chargeholestate}) and (\ref{cformtr}) respectively, we find that the charge of a soliton is

\be |L\Delta\tilde{\rho}_{\theta}(0)|=\frac{\pi}{2(\pi-u)}=q. \ee
The energy of the soliton is given by

\be E_{\theta}=-Lm_0\int_{-\Lambda}^{\Lambda}d\alpha \; \cosh\alpha \;\Delta\rho_{\theta}(\alpha).
\ee
This can be expressed in terms of the Fourier transforms of $\cosh\alpha$ and $\Delta\rho_{\theta}(\alpha)$ as

 \be E_{\theta}= \frac{m_0 }{4\pi}\int_{-\infty}^{\infty}d\omega\;\frac{e^{\Lambda}}{2}\left(\frac{e^{i\Lambda \omega}}{1+i\omega}+ \frac{e^{-i\Lambda \omega}}{1-i\omega}\right)\; \frac{\sinh(\pi\omega)\cos(\theta\omega)}{\sinh((\pi-u)\omega)\cosh(u\omega)},\label{enholeeq}
 \ee
where (\ref{FTforms}) and (\ref{denhole}) were used. In the limit $\Lambda\rightarrow\infty$, the poles close to the $x-$ axis arising from the term $\cosh(u\omega)$ contribute, whereas the contribution from the other terms vanishes exponentially \cite{Thacker}. The integration is performed by closing the contour in the upper and lower half planes for the first and the second terms of $\tilde{c}(\omega)$ respectively. We obtain 

\be\label{etheta} E_{\theta}= m \cosh(\gamma\theta), \;\;\; m= \frac{m_0\gamma}{\pi(\gamma-1)}\tan(\pi\gamma)\;e^{\Lambda(1-\gamma)}, \;\;\; \gamma=\frac{\pi}{2u},
\ee
 where $m$ is the renormalized mass. Instead of considering the Bethe equations corresponding to charge conjugated fermions, by considering the Bethe equations of the original fermions and adding a hole one obtains anti-solitons which carry charge $-q$, which is exactly opposite to that of the solitons. The energy of the anti-solitons is exactly same as that of the solitons (\ref{etheta}). Hence we see that the first excited state above the ground state $\ket{0}$ is doubly degenerate with normal ordered charge $\mathcal{N}=\pm 1$, where $\pm$ correspond to the soliton and anti-soliton respectively.

\subsection{Mid-gap phase: $A_1$}

The mid-gap phase is divided into four sub-phases $A_i, i=1,2,3,4$ depending on the values of the boundary fields $\phi_{L,R}$.  For $m_0>0$, they correspond to the following domains of the boundaries fields $0<\phi_{L,R}<u$, $(-u<\phi_L<0,0<\phi_R<u),  (-u<\phi_L<0,-u<\phi_R<0),(0<\phi_L<u,-u<\phi_R<0)$ respectively. For the case of $m_0<0$, as mentioned before, both the boundary fields are shifted by $\phi_{L,R}\rightarrow \phi_{L,R}+i\pi$. Below we shall provide an explicit solution to the Bethe equations (\ref{BEMT}) in the $A_1$ phase and obtain the low energy spectrum.

\subsubsection{The state $|-1\rangle$}
Let us consider the Bethe equations in the mid-gap (\ref{BEMT}) corresponding to the original fermions. Let us consider the case where $m_0>0$, such that the boundary fields lie in the range $\phi_{L,R}\in (0,u)$. As mentioned before, from the relation between the Bethe roots $\beta_j$ and the energy (\ref{eneq}), we can infer that the roots lying on the line $\beta_j=\alpha_j+i\pi$ have lower energy. Since we are interested in the low energy spectrum, we consider the state which has all the roots lying on this line.  By making the transformation $\beta_j\rightarrow\alpha_j+i\pi$ and applying logarithm to (\ref{BEMT}), we obtain the logarithmic form of the Bethe equations in the $A_1$ phase
\bea \nonumber-2im_0L\sinh\alpha_i=\ln\left(\frac{\sinh\left(\frac{1}{2}(\alpha_i+i\phi_L+iu)\right)}{\sinh\left(\frac{1}{2}(\alpha_i-i\phi_L-iu)\right)}\right)+\ln\left(\frac{\sinh\left(\frac{1}{2}(\alpha_i+i\phi_R+iu)\right)}{\sinh\left(\frac{1}{2}(\alpha_i-i\phi_R-iu)\right)}\right)\\-2i\pi n_i+\ln\left(\frac{\sinh\left(\frac{1}{2}(\alpha_i-iu)\right)}{\sinh\left(\frac{1}{2}(\alpha_i+iu)\right)}\right)-\sum_{\sigma=\pm}\sum_j\ln\left(\frac{\sinh\left(\frac{1}{2}(\alpha_i-\sigma\alpha_j-2iu)\right)}{\sinh\left(\frac{1}{2}(\alpha_i-\sigma\alpha_j+2iu)\right)}\right).\label{BEsup}\eea
Using the same notation as before, $\varphi(x,y)=i\ln\left(\frac{(\sinh(x+iy)/2)}{(\sinh(x-iy)/2)}\right)$, the above equation can be written as

\bea 2m_0L\sinh\alpha_i=\varphi(\alpha_i,\phi_R+u)+\varphi(\alpha_i,\phi_L+u)-\sum_{\sigma=\pm}\sum_{j}\varphi(\alpha_i-i+\sigma\alpha_j,2u)-\varphi(2\alpha_i,2u)+2\pi n_i.\label{belg}\eea
The above equation can be solved by following the same procedure as in the trivial phase. Let us denote the density distribution of the resulting state by $\rho_{\ket{-1}}(\alpha)$. The reason for this notation will become evident soon. We obtain

\bea \nonumber2m_0\cosh\alpha-\frac{1}{L}\varphi^{\prime}(\alpha,\phi_R+u)-\frac{1}{L}\varphi^{\prime}(\alpha,\phi_L+u)-4\pi\rho_{\ket{-1}}(\alpha)-\frac{1}{L}\delta(\alpha)+\frac{1}{L}\varphi^{\prime}(\alpha,u)+\frac{1}{L}\varphi^{\prime}(\alpha,u+i\pi)\\=\sum_{\sigma}\sum_j\frac{1}{L}\varphi^{\prime}(\alpha+\sigma\gamma_j,2u)=\sum_{\sigma}\int\varphi^{\prime}(\alpha+\sigma\gamma,2u)\rho_{\ket{-1}}(\gamma)d\gamma,
\label{gsden}\eea
where \be\varphi^{\prime}(x,y)=\partial_x\varphi(x,y)=\frac{\sin y}{\cosh x-\cos y},\;\;\;\varphi^{\prime}(x,y+i\pi)=\partial_x\varphi(x,y+i\pi)=-\frac{\sin y}{\cosh x+\cos y}.\ee

The above equation (\ref{gsden}) can be solved by applying Fourier transform just as in the previous case. Following the same procedure, we obtain

\bea \label{dentop-}\tilde{\rho}_{\ket{-1}}(\omega)=\tilde{\rho}_{Bulk}(\omega)+\tilde{\rho}_{Boundary}(\omega)+\tilde{\rho}^{top}_{Boundary fields}(\omega),\eea
where $\tilde{\rho}_{Bulk}(\omega)$ and $\tilde{\rho}_{Boundary}(\omega)$ are same as in the trivial phase (\ref{denbulk}), (\ref{denb}), whereas 

\bea\label{denbtop-}\tilde{\rho}^{top}_{Boundary fields}(\omega)=-\frac{1}{2L}\sum_{a=L,R}\frac{\sinh((\pi-(\phi_a+u))\omega)}{\sinh(\pi\omega)+\sinh((\pi-2u)\omega)}.\eea
The charge of the state obtained above by filling the roots continuously on the $i\pi$ line, which is described by the density distribution $\rho_{\ket{-1}}(\alpha)$ is 

\be \label{tcharge-1A1} L\int_{-\Lambda}^{\Lambda} d\alpha\;\rho_{\ket{-1}}(\alpha)= L\tilde{\rho}_{\ket{-1}}(0)=\frac{m_0 L e^{\Lambda}}{2\pi}\;\frac{\pi}{2(\pi-u)}-\frac{\pi}{2(\pi-u)}+\frac{\phi_L+\phi_R}{4(\pi-u)}.
\ee
The first term which is divergent in the limit $\Lambda\rightarrow\infty$ is the charge associated with the bulk. The second and the third terms are due to the charge associated with the boundaries. This can be expressed as

\be\label{cform-1} L\tilde{\rho}_{\ket{-1}}(0)= q\left(\frac{m_0 L e^{\Lambda}}{2\pi}-1+\frac{\phi_L+\phi_R}{2\pi}\right), \;\; \;q=\frac{\pi}{2(\pi-u)}.\ee
Defining the normal ordered charge of the state as before, we have
\be\mathcal{N}=-1+\frac{\phi_L+\phi_R}{2\pi}.\label{charge-1}\ee
Absorbing the boundary condition into the normal ordered charge as before (\ref{Nprime}), we see that the charge of the state described by the density distribution $\rho_{\ket{-1}}(\alpha)$ (\ref{dentop-}) is
\be\mathcal{N}=-1.\label{charge-1state}
\ee
This allows us to label this state as $\ket{-1}$.

\vspace{3mm}

\subsubsection{The state $|1\rangle$}

In the previous section we have obtained the state with negative normal ordered charge. To obtain the states with positive normal ordered charge, we need to invoke charge conjugation defined above (\ref{apcc}). Consider the state with all Bethe roots lying on the $i\pi$ line just as in the case of the state $\ket{-1}$ discussed above. By making the transformation $\beta_j\rightarrow\alpha_j+i\pi$ and applying logarithm to (\ref{BEMTCC}), we obtain the logarithmic form of the Bethe equations 
\bea \nonumber-2im_0L\sinh\alpha_i=\ln\left(\frac{\sin\left(\frac{1}{2}(\alpha_i-i\phi_L+iu)\right)}{\sinh\left(\frac{1}{2}(\alpha_i+i\phi_L-iu)\right)}\right)+\ln\left(\frac{\sinh\left(\frac{1}{2}(\alpha_i-i\phi_R+iu)\right)}{\sinh\left(\frac{1}{2}(\alpha_i+i\phi_R-iu)\right)}\right)\\-2i\pi n_i+\ln\left(\frac{\sinh\left(\frac{1}{2}(\alpha_i-iu)\right)}{\sinh\left(\frac{1}{2}(\alpha_i+iu)\right)}\right)-\sum_{\sigma=\pm}\sum_j\ln\left(\frac{\sinh\left(\frac{1}{2}(\alpha_i-\sigma\alpha_j-2iu)\right)}{\sinh\left(\frac{1}{2}(\alpha_i-\sigma\alpha_j+2iu)\right)}\right).\label{logBE2}\eea
Following the same procedure as above (\ref{defden}), and defining the density distribution $\rho_{\ket{1}}(\alpha)$ associated with the roots, we obtain

\bea \nonumber2m_0\cosh\alpha-\frac{1}{L}\varphi^{\prime}(\alpha,-\phi_R+u)-\frac{1}{L}\varphi^{\prime}(\alpha,-\phi_L+u)-4\pi\rho_{\ket{+1}}(\alpha)-\frac{1}{L}\delta(\alpha)+\frac{1}{L}\varphi^{\prime}(\alpha,u)+\frac{1}{L}\varphi^{\prime}(\alpha,u+i\pi)\\=\sum_{\sigma}\sum_j\frac{1}{L}\varphi^{\prime}(\alpha+\sigma\gamma_j,2u)=\sum_{\sigma}\int\varphi^{\prime}(\alpha+\sigma\gamma,2u)\rho_{\ket{1}}(\gamma)d\gamma,
\label{gsden2}\eea
This equation can be solved by applying Fourier transform as before. We obtain

\bea \label{dentop+}\tilde{\rho}_{\ket{1}}(\omega)=\tilde{\rho}_{Bulk}(\omega)+\tilde{\rho}_{Boundary}(\omega)+\tilde{\rho}^{top'}_{Boundary fields}(\omega),\eea
where $\rho_{Bulk}(\omega)$, $\tilde{\rho}_{Boundary}(\omega)$ are same as that in the previous case (\ref{denbulk}),\(\ref{denb})\) and

\bea \label{denbtop+}\tilde{\rho}^{top'}_{Boundary fields}(\omega)=-\frac{1}{2L}\sum_{a=L,R}\frac{\sinh((\pi-(u-\phi_a))\omega)}{\sinh(\pi\omega)+\sinh((\pi-2u)\omega)}.\eea
Hence the charge of the state described by the density distribution $\rho_{\ket{1}}(\alpha)$ obtained above is

\be \label{tcharge+1A1} -L\int_{-\Lambda}^{\Lambda} d\alpha\;\rho_{\ket{1}}(\alpha)= -L\tilde{\rho}_{\ket{1}}(0)=-\left(\frac{m_0 L e^{\Lambda}}{2\pi}\;\frac{\pi}{2(\pi-u)}-\frac{\pi}{2(\pi-u)}-\frac{\phi_L+\phi_R}{4(\pi-u)}\right).
\ee
Hence, the normal ordered charge of a state associated with charge conjugated fermions (up to an unimportant constant) is

\be\mathcal{N}= -\frac{1}{q}\left\{\tilde{\rho}(0)-\tilde{\rho}_{Bulk}(0) \right\}.\ee
Using this we obtain the normal ordered charge associated with the state described by the root distribution $\rho_{\ket{1}}(\alpha)$ as 

\bea \mathcal{N}=1+\frac{\phi_L+\phi_R}{2\pi}.\label{charge1}\eea

Notice that the latter result could have been obtained from (\ref{charge-1}) 
by charge conjugation, i.e through  simultaneous changes:
$\mathcal{N} \rightarrow -\mathcal{N}$ and $\phi_{L,R} \rightarrow -\phi_{L,R} $
as it should. In a similar way as before (\ref{Nprime}), absorbing the boundary fields into the normal ordering, we have 

\be \mathcal{N}=1.
\ee

Hence, we label this state as $\ket{1}$.

\subsubsection{The states $|0\rangle_{\cal L}$ and $|0\rangle_{\cal R}$}

The Bethe equations corresponding to the charge conjugated fermions (\ref{BEMTCC}) have two boundary string solutions: $\pm i (u-\phi_b), \; b=L,R$. In the present case, the boundary strings fall under the category named `close' or `short' boundary strings when $|\phi_{L,R}|<u$ \cite{YSRKondo,Parmeshthesis}. These boundary string solutions correspond to exponentially localized bound states \cite{skorik}. Consider adding one boundary string, either $b=L$ or $b=R$, to the state described by the density distribution $\rho_{\ket{1}}(\alpha)$. Labelling the density distribution associated with the resulting state as $\rho_{\phi_b}(\alpha)$, we have

\bea \nonumber2m_0\cosh\alpha-\frac{1}{L}\varphi^{\prime}(\alpha,-\phi_R+u)-\frac{1}{L}\varphi^{\prime}(\alpha,-\phi_L+u)-4\pi\rho_{\phi_b}(\alpha)-\frac{1}{L}\delta(\alpha)+\frac{1}{L}\varphi^{\prime}(\alpha,u)+\frac{1}{L}\varphi^{\prime}(\alpha,u+i\pi)\\=\frac{1}{L}\varphi^{\prime}(\alpha,-\phi_b+3u)+\frac{1}{L}\varphi^{\prime}(\alpha,u+\phi_b)+\sum_{\sigma}\int\varphi^{\prime}(\alpha+\sigma\gamma,2u)\rho_{\phi_b}(\gamma)d\gamma.
\label{gsdenbs}\eea
Following the same procedure as above, we obtain

\be \label{denphi}\tilde{\rho}_{\phi_b}(\omega)=\tilde{\rho}_{\ket{1}}(\omega)+\Delta\tilde{\rho}_{\phi_b}(\omega), \;\;\; \Delta\tilde{\rho}_{\phi_b}(\omega)=-\frac{1}{L}\;\frac{\sinh((\pi-2u)\omega)\cosh((u-\phi_b)\omega)}{\sinh(\pi\omega)+\sinh((\pi-2u)\omega)},\ee
where $\tilde{\rho}_{\ket{1}}(\omega)$ is given by (\ref{dentop+},\ref{denbtop+}). We see that by adding the boundary string $\pm i (u-\phi_b)$, $b=L,R$, to the state $\ket{1}$ results in the change $\Delta\tilde{\rho}_{\phi_b}(\alpha)$ to the root distribution $\rho_{\ket{1}}(\alpha)$   due to the back flow effect. This results in a new density distribution $\rho_{\phi_b}(\alpha)$ for the Bethe roots lying on the $i\pi$ line. Hence, we obtain a new state which is described by the set of Bethe roots which in addition to roots on the $i\pi$ line with the density distribution $\rho_{\phi_b}(\alpha)$, includes the boundary string  $\pm i (u-\phi_b)$, $b=L,R$. The total charge of this state is

\be 1+L\tilde{\rho}_{\phi_b}(0)= \frac{m_0 L e^{\Lambda}}{2\pi}\;\frac{\pi}{2(\pi-u)}-\frac{\phi_L+\phi_R}{4(\pi-u)}.\ee
The first term `$1$' on the left side of the above equation corresponds to the boundary string and the second term $L\tilde{\rho}_{\phi_b}(0)$ corresponds to the roots on the $i\pi$ line. Hence the normal ordered charge of this state is 
\be
\mathcal{N}=\frac{\phi_L+\phi_R}{2\pi}
\label{charge0LRa1}.
\ee
 Comparing the charge of this state with the charge of the state described by the density distribution $\rho_{\ket{1}}(\alpha)$, we find that the net charge of the boundary string $\pm i(-\phi_b+u)$ is exactly equal and opposite to that of a soliton. Depending on whether the boundary string $\pm i(u-\phi_L)$ or $\pm i(u-\phi_R)$ is added to the state $|1\rangle$, we obtain two states with same charge given by (\ref{charge0LRa1}).
\vspace{3mm}

To compare the energies of the two states described above, we need to calculate the energy of the boundary string $\pm i(u-\phi_b)$, $b=L,R $ (with respect to that of the $|1\rangle$ state). This is given by

\be E_{\phi_b}=-m_0\cosh(i(-\phi_b+u))-m_0L\int_{-\Lambda}^{\Lambda}d\alpha\;\cosh\alpha \;\Delta\rho_{\phi_b}(\alpha).
\ee
This can be written as

\be E_{\phi_b} =-m_0L\int_{-\Lambda}^{\Lambda}d\alpha \; \cosh\alpha \; \left(\frac{1}{2L}\sum_{\sigma=\pm}\delta(\alpha+\sigma i(-\phi_b+u))+\Delta\rho_{\phi_b}(\alpha)\right).
\ee
This can be expressed in terms of the Fourier transforms of $\cosh\alpha$ and term in the parenthesis. We have

\be E_{\phi_b}=-\frac{m_0}{4\pi}\int_{-\infty}^{\infty}d\omega \;\frac{e^{\Lambda}}{2}\left(\frac{e^{i\Lambda \omega}}{1+i\omega}+ \frac{e^{-i\Lambda \omega}}{1-i\omega}\right)\frac{\sinh(\pi\omega)\cosh((u-\phi_b)\omega)}{\sinh((\pi-u)\omega)\cosh(u\omega)}.\label{enbseq}
\ee
Comparing this equation with that of (\ref{enholeeq}), we see that it can be evaluated in the same way. We obtain

\be\label{ebs} E_{\phi_b}=-m\sin(\gamma\phi_b)\equiv -m_{b}, 
\ee
where $b=L,R$.
Hence we find that the energy of the boundary string takes similar form as that of the soliton (\ref{etheta}). We see that by adding the boundary string $\pm i(-\phi_b+u)$, $b=L,R$ to the state $\ket{1}$, we obtain a state which has energy $E_{\ket{1}}-m_b$, $b=L,R$ respectively, where $0<m_b<m$.

\vspace{3mm}

Now let us calculate the energy difference between the states $\ket{\pm1}$ described above. This is given by
\bea E_{\ket{1}}-E_{\ket{-1}}= -L\int_{-\Lambda}^{\Lambda} d\alpha \; m_0\cosh\alpha \;(\rho_{\ket{1}}(\alpha)-\rho_{\ket{-1}}(\alpha))=-\frac{m_0L}{2\pi} \int_{-\infty}^{\infty}d\omega \tilde{c}(\omega)(\rho_{\ket{1}}(\omega)-\rho_{\ket{-1}}(\omega)).
\eea
 Using (\ref{FTforms},\ref{dentop+},\ref{denbtop+},\ref{dentop-},\ref{denbtop-}), and following the same procedure as above we obtain
\be \label{diffen}E_{\ket{1}}-E_{\ket{-1}}= m_L+m_R,
\ee 
where $m_{L,R}$ are given by (\ref{ebs}). We thus finally end up with the following picture: Adding a right boundary string to the state $\ket{1}$  gives a state with energy $E_{\ket{-1}}+ m_L$ while adding a left boundary string to the state $\ket{1}$ gives a state which has energy $E_{\ket{-1}}+ m_R$. Since $m_{L,R} >0$ in the $A_1$ phase,  we see that $\ket{-1}$ is the ground state. We interpret that the states which contain the boundary strings $\pm i(u-\phi_L)$, $\pm i(u-\phi_R)$ which have energies $m_R$ and $m_L$ respectively above the ground state $\ket{-1}$, host localized bound states at the right and left boundaries respectively. The charge of these two states is given by (\ref{charge0LRa1}). Absorbing the boundary fields into the normal ordering as before (\ref{Nprime}), we see that the normal ordered charge of the states which contain the boundary string on top of the state $\ket{1}$ is

\be\mathcal{N}=0.
\ee
Hence, we label these two states
by
\be
\ket{0}_{\cal R}\; {\rm and}\; \ket{0}_{\cal L}.
\ee
The subscript denotes the edge where the bound state is present. Finally, the state $\ket{1}$  is seen as  hosting two bound states, one at each edge, with  energy $m_L+m_R$ above the ground state $\ket{-1}$.   
  
\vspace{5mm}

Note that just like the Bethe equations corresponding to the charge conjugated fermions, the Bethe equations of the original fermions (\ref{BEMT}) have two boundary strings $\pm i (u+\phi_{b})$, $b=L,R$. One may naturally question whether the states $\ket{0}_{\cal L}$, $\ket{0}_{\cal R}$ can be obtained by adding these boundary strings to the state $\ket{-1}$. It turns out that it is not possible. This is due to the following reason: As mentioned before, the charge of any state cannot be greater than the maximum charge (\ref{MaxC}). In the mid-gap phase, due to the effect of the boundaries, the difference of the charge of the state $\ket{1}$ (\ref{tcharge+1A1}) and that of the maximum charge (\ref{MaxC}) is large enough such that a close (short) boundary string could be added to the state $\ket{1}$ yielding a state whose total charge is less than the  maximum charge (\ref{MaxC}). For the state $\ket{-1}$, the difference in the charge of the state and that of the maximum charge is not large enough to add a close (short) boundary string, since the charge of the resulting state would exceed the maximum charge (\ref{MaxC}). However note that when both the boundary fields $\phi_{L,R}=0$ in the Bethe equations (\ref{BEMT}), (\ref{BEMTCC}), the charge of the two states $\ket{\pm1}$ is exactly the same and it is possible to obtain both the states $\ket{0}_{\cal L}$ and $\ket{0}_{\cal R}$ by adding respective boundary strings to either $\ket{1}$ state or $\ket{-1}$ state.

  \subsubsection{Summary Phase $A_1$:} 
  
  1) There exists four states $\ket{\pm1}, \ket{0}_{L}, \ket{0}_{R}$ whose energy is below the mass gap $m$. 
  
  2) Both the states $\ket{\pm1}$ contain Bethe roots which lie on the $i\pi$ line. The energy difference between these states is $m_L+m_R$. 
  
  3) Since $m_{L,R}>0$,  the ground state in the phase $A_1$ is $\ket{-1}$. The state $\ket{1}$ contains two bound states at each edge and has energy $m_L+m_R$ above the ground state. 

4) The states $\ket{0}_{R},\ket{0}_{L}$ are obtained by adding the boundary strings $\pm i(u-\phi_{L}), \pm i(u-\phi_R)$ to the state $\ket{1}$ respectively. 

5) The boundary strings $\pm i(u-\phi_{L,R})$ have energies $-m_{L,R}$ respectively, and hence the state $\ket{0}_L$ has energy $-m_L$ with respect to the state $\ket{1}$ or equivalently, it has energy $m_R$ with respect to the state $\ket{-1}$. Similarly, the state $\ket{0}_R$ has energy $-m_R$ with respect to the state $\ket{1}$ or equivalently, it has energy $m_L$ with respect to the state $\ket{-1}$.

6) Hence we interpret that by the state $\ket{0}_L\;(\ket{0}_R)$ contains a bound state with energy $m_L\; (m_R)$ at the left (right) edge on top of the ground state $\ket{-1}$.

7) Comparing the energies of the two states $\ket{0}_{\cal L}$ and $\ket{0}_{\cal R}$, we find that they are degenerate on the line $\phi_L=\phi_R$.

\subsection{Mid-gap phase: $A_2$ }
In this section we present the solution to the Bethe equations (\ref{BEMT}) in the $A_2$ phase. In this phase the boundary fields $\phi_{L,R}$ take values in the following range: $-u<\phi_L<0$, $0<\phi_R<u$. This phase can be further split into two sub-regions depending on the absolute values of the boundary fields. Let us first consider the case where $\phi_R>|\phi_L|$. 
\subsubsection{The state \ket{-1} ($\phi_R>|\phi_L|$)}
It is easier to work with the notation where the boundary parameters in the Bethe equations are positive. In this region since $\phi_L<0$, we use the transformation $\phi_L\rightarrow-\phi^{\prime}_L$, where $\phi^{\prime}_L\equiv |\phi_L|$, to express the Bethe equations in terms of the absolute values of $\phi_{L,R}$. 
The Bethe equations corresponding to the original fermions are
\bea\nonumber
e^{2im_0L\sinh\beta_i}=\frac{\cosh\left(\frac{1}{2}(\beta_i-i\phi^{\prime}_L+iu)\right)}{\cosh\left(\frac{1}{2}(\beta_i+i\phi^{\prime}_L-iu)\right)}\frac{\cosh\left(\frac{1}{2}(\beta_i+i\phi_R+iu)\right)}{\cosh\left(\frac{1}{2}(\beta_i-i\phi_R-iu)\right)}\prod_{i\neq j,j=1}^N\frac{\sinh\left(\frac{1}{2}(\beta_i-\beta_j+2iu)\right)}{\sinh\left(\frac{1}{2}(\beta_i-\beta_j-2iu)\right)}\frac{\sinh\left(\frac{1}{2}(\beta_i+\beta_j+2iu)\right)}{\sinh\left(\frac{1}{2}(\beta_i+\beta_j-2iu)\right)}.\\\label{BEMTA2}\eea
Consider the state with all Bethe roots lying on the $i\pi$ line just as in the case of the state $\ket{-1}$ in the $A_1$ phase. By making the transformation $\beta_j\rightarrow\alpha_j+i\pi$ and applying logarithm to (\ref{BEMTA2}), we obtain the logarithmic form of the Bethe equations

\bea \nonumber-2im_0L\sinh\alpha_i=\ln\left(\frac{\sin\left(\frac{1}{2}(\alpha_i-i\phi^{\prime}_L+iu)\right)}{\sinh\left(\frac{1}{2}(\alpha_i+i\phi^{\prime}_L-iu)\right)}\right)+\ln\left(\frac{\sinh\left(\frac{1}{2}(\alpha_i+i\phi_R+iu)\right)}{\sinh\left(\frac{1}{2}(\alpha_i-i\phi_R-iu)\right)}\right)\\-2i\pi n_i+\ln\left(\frac{\sinh\left(\frac{1}{2}(\alpha_i-iu)\right)}{\sinh\left(\frac{1}{2}(\alpha_i+iu)\right)}\right)-\sum_{\sigma=\pm}\sum_j\ln\left(\frac{\sinh\left(\frac{1}{2}(\alpha_i-\sigma\alpha_j-2iu)\right)}{\sinh\left(\frac{1}{2}(\alpha_i-\sigma\alpha_j+2iu)\right)}\right).\label{BEsupA2-1}\eea
By following the same procedure as above (\ref{defden}) and define the density distribution $\rho_{\ket{-1}}(\alpha)$ associated with the roots, we have

\bea \nonumber2m_0\cosh\alpha-\frac{1}{L}\varphi^{\prime}(\alpha,\phi_R+u)-\frac{1}{L}\varphi^{\prime}(\alpha,-\phi^{\prime}_L+u)-4\pi\rho_{\ket{-1}}(\alpha)-\frac{1}{L}\delta(\alpha)+\frac{1}{L}\varphi^{\prime}(\alpha,u)+\frac{1}{L}\varphi^{\prime}(\alpha,u+i\pi)\\=\sum_{\sigma}\sum_j\frac{1}{L}\varphi^{\prime}(\alpha+\sigma\gamma_j,2u)=\sum_{\sigma}\int\varphi^{\prime}(\alpha+\sigma\gamma,2u)\rho_{\ket{-1}}(\gamma)d\gamma,
\label{gsdena2-1}\eea
Just as before, this equation can be solved by Fourier transform. We obtain

\bea \label{dena2-1}\tilde{\rho}_{\ket{-1}}(\omega)=\tilde{\rho}_{Bulk}(\omega)+\tilde{\rho}_{Boundary}(\omega)+\tilde{\rho}^{top}_{Boundary fields}(\omega),\eea
where $\tilde{\rho}_{Bulk}(\omega)$, $\tilde{\rho}_{Boundary}(\omega)$ take the same form as (\ref{denbulk}), (\ref{denb}) and

\bea \tilde{\rho}^{top}_{Boundary fields}(\omega)=-\frac{1}{2L}\frac{\sinh((\pi-(\phi_R+u))\omega)}{\sinh(\pi\omega)+\sinh((\pi-2u)\omega)}-\frac{1}{2L}\frac{\sinh((\pi-(u-\phi^{\prime}_L))\omega)}{\sinh(\pi\omega)+\sinh((\pi-2u)\omega)}\label{denba2-1}.\eea
The charge of this state is 

\be \label{charge-1A2} L\tilde{\rho}_{\ket{-1}}(0)=\frac{m_0 L e^{\Lambda}}{2\pi}\;\frac{\pi}{2(\pi-u)}-\frac{\pi}{2(\pi-u)}+\frac{\phi_R-\phi^{\prime}_L}{4(\pi-u)}.
\ee
Hence, as expected, the normal ordered charge of the state is 

\be \mathcal{N}=-1+\frac{\phi_R-\phi^{\prime}_L}{2\pi}.\ee
Reverting back to the original boundary variables, we have

\be \mathcal{N}=-1+\frac{\phi_R+\phi_L}{2\pi}.\ee
Just as before (\ref{Nprime}), absorbing the boundary fields into the definition of normal ordering, we have
\be\mathcal{N}=-1.
\ee
Hence we label this state $\ket{-1}$.

\subsubsection{The state \ket{1} ($\phi_R>|\phi_L|$)}
To obtain states with positive charge, just as in the case of $A_1$ phase, we need to work with the charge conjugated fermions. The Bethe equations take the form given in (\ref{BEMTA2}) with the transformation $\phi_{R},\phi^{\prime}_L\rightarrow -\phi_{R},-\phi^{\prime}_L$ respectively. The Bethe equations corresponding to the charge conjugated fermions are 

\bea\nonumber
e^{2im_0L\sinh\beta_i}=\frac{\cosh\left(\frac{1}{2}(\beta_i+i\phi^{\prime}_L+iu)\right)}{\cosh\left(\frac{1}{2}(\beta_i-i\phi^{\prime}_L-iu)\right)}\frac{\cosh\left(\frac{1}{2}(\beta_i-i\phi_R+iu)\right)}{\cosh\left(\frac{1}{2}(\beta_i+i\phi_R-iu)\right)}\prod_{i\neq j,j=1}^N\frac{\sinh\left(\frac{1}{2}(\beta_i-\beta_j+2iu)\right)}{\sinh\left(\frac{1}{2}(\beta_i-\beta_j-2iu)\right)}\frac{\sinh\left(\frac{1}{2}(\beta_i+\beta_j+2iu)\right)}{\sinh\left(\frac{1}{2}(\beta_i+\beta_j-2iu)\right)}.\\\label{BEMTA2+1}\eea
Just as before, considering the state with all roots lying on the $i\pi$ line and making the transformation $\beta_j\rightarrow\alpha_j+i\pi$ and applying logarithm to (\ref{BEMTA2+1}), we obtain the logarithmic form of the Bethe equations

\bea \nonumber-2im_0L\sinh\alpha_i=\ln\left(\frac{\sin\left(\frac{1}{2}(\alpha_i+i\phi^{\prime}_L+iu)\right)}{\sinh\left(\frac{1}{2}(\alpha_i-i\phi^{\prime}_L-iu)\right)}\right)+\ln\left(\frac{\sinh\left(\frac{1}{2}(\alpha_i-i\phi_R+iu)\right)}{\sinh\left(\frac{1}{2}(\alpha_i+i\phi_R-iu)\right)}\right)\\-2i\pi n_i+\ln\left(\frac{\sinh\left(\frac{1}{2}(\alpha_i-iu)\right)}{\sinh\left(\frac{1}{2}(\alpha_i+iu)\right)}\right)-\sum_{\sigma=\pm}\sum_j\ln\left(\frac{\sinh\left(\frac{1}{2}(\alpha_i-\sigma\alpha_j-2iu)\right)}{\sinh\left(\frac{1}{2}(\alpha_i-\sigma\alpha_j+2iu)\right)}\right).\label{BEsupA2+1}\eea
By following the same procedure as above (\ref{defden}) and define the density distribution $\rho_{\ket{1}}(\alpha)$ associated with the roots, we have

\bea \nonumber2m_0\cosh\alpha-\frac{1}{L}\varphi^{\prime}(\alpha,-\phi_R+u)-\frac{1}{L}\varphi^{\prime}(\alpha,\phi^{\prime}_L+u)-4\pi\rho_{\ket{1}}(\alpha)-\frac{1}{L}\delta(\alpha)+\frac{1}{L}\varphi^{\prime}(\alpha,u)+\frac{1}{L}\varphi^{\prime}(\alpha,u+i\pi)\\=\sum_{\sigma}\sum_j\frac{1}{L}\varphi^{\prime}(\alpha+\sigma\gamma_j,2u)=\sum_{\sigma}\int\varphi^{\prime}(\alpha+\sigma\gamma,2u)\rho_{\ket{1}}(\gamma)d\gamma,
\label{gsdena2+1}\eea
Just as before, this equation can be solved by Fourier transform. We obtain

\bea \label{denforma2+1}\tilde{\rho}_{\ket{1}}(\omega)=\tilde{\rho}_{Bulk}(\omega)+\tilde{\rho}_{Boundary}(\omega)+\tilde{\rho}^{top'}_{Boundary fields}(\omega),\eea
where $\rho_{Bulk}(\omega)$, $\tilde{\rho}_{Boundary}(\omega)$ are the same as in the previous case (\ref{denbulk}), (\ref{denb}) and $\tilde{\rho}^{top'}_{Boundary fields}(\omega)$ takes the form given in (\ref{denba2-1}) with the change $\phi^{\prime}_L\rightarrow -\phi^{\prime}_L$, $\phi_R\rightarrow-\phi_R$.  Explicitly we have

\bea\nonumber \tilde{\rho}^{top'}_{Boundary fields}(\omega)=-\frac{1}{2L}\frac{\sinh((\pi-(u-\phi_R))\omega)}{\sinh(\pi\omega)+\sinh((\pi-2u)\omega)}-\frac{1}{2L}\frac{\sinh((\pi-(\phi^{\prime}_L-u))\omega)}{\sinh(\pi\omega)+\sinh((\pi-2u)\omega)}\label{denba2+1}.\eea
The charge of this state is 

\be \label{charge+1A2} L\tilde{\rho}_{\ket{1}}(0)=\frac{m_0 L e^{\Lambda}}{2\pi}\;\frac{\pi}{2(\pi-u)}-\frac{\pi}{2(\pi-u)}-\frac{\phi_R-\phi^{\prime}_L}{4(\pi-u)}.
\ee
As we expect, we find that the normal ordered charge of the state is 

\be\mathcal{N}=1+\frac{\phi_R-\phi^{\prime}_L}{2\pi}.
\ee
Reverting back to the original boundary variables, we have

\be \mathcal{N}=1+\frac{\phi_R+\phi_L}{2\pi}.\ee
Absorbing the boundary fields into the normal ordered charge as before (\ref{Nprime}), we have 

\be \mathcal{N}=1,
\ee
and hence we label this state $\ket{1}$.

\subsubsection{The states $|0\rangle_{\cal L}$ and $|0\rangle_{\cal R}$ \;($\phi_R>|\phi_L|$)}
The Bethe equations corresponding to the charge conjugated fermions in the $A_2$ phase have two boundary string solutions: $\pm i (u-\phi_R)$, $\pm i (u+\phi^{\prime}_L)$. Recall that $\phi^{\prime}_L=|\phi_L|$. Note that unlike in the $A_1$ phase, the boundary strings corresponding to the left and right boundaries have a slightly different form since the boundary fields $\phi_{L,R}$ have opposite signs. Nevertheless, just as in the $A_1$ phase, both these boundary strings fall into the close (short) boundary strings category. Consider adding the right boundary string $\pm i (u-\phi_R)$ to the state $\ket{1}$ described above. Labelling the density distribution of the resulting state as $\rho_{\phi_R}(\alpha)$, we have

\bea \nonumber2m_0\cosh\alpha-\frac{1}{L}\varphi^{\prime}(\alpha,-\phi_R+u)-\frac{1}{L}\varphi^{\prime}(\alpha,\phi^{\prime}_L+u)-4\pi\rho_{\phi_R}(\alpha)-\frac{1}{L}\delta(\alpha)+\frac{1}{L}\varphi^{\prime}(\alpha,u)+\frac{1}{L}\varphi^{\prime}(\alpha,u+i\pi)\\=\frac{1}{L}\varphi^{\prime}(\alpha,-\phi_R+3u)+\frac{1}{L}\varphi^{\prime}(\alpha,u+\phi_R)+\sum_{\sigma}\int\varphi^{\prime}(\alpha+\sigma\gamma,2u)\rho_{\phi_R}(\gamma)d\gamma.
\label{gsdenbsa2r}\eea
Following the same procedure as in the $A_1$ phase, we obtain

\be \tilde{\rho}_{\phi_R}(\omega)=\tilde{\rho}_{\ket{1}}(\omega)+\Delta\tilde{\rho}_{\phi_R}(\omega), \;\;\; \Delta\tilde{\rho}_{\phi_R}(\omega)=-\frac{1}{L}\;\frac{\sinh((\pi-2u)\omega)\cosh((u-\phi_R)\omega)}{\sinh(\pi\omega)+\sinh((\pi-2u)\omega)},\ee
where $\tilde{\rho}_{\ket{1}}(\omega)$ is given by (\ref{denforma2+1}), (\ref{denba2+1}). The total charge of this state is

\be 1+L\tilde{\rho}_{\phi_R}(0)=\frac{m_0 L e^{\Lambda}}{2\pi}\;\frac{\pi}{2(\pi-u)}-\frac{\phi_R-\phi^{\prime}_L}{4(\pi-u)}.\ee
Since $\phi_R> \phi^{\prime}_L (|\phi_L|)$, the total charge of the above state is less than the maximum allowed charge (\ref{MaxC}), and hence this is a valid state. By following the same procedure, we find that the normal ordered charge of this state is
\be
\mathcal{N}=\frac{\phi_R-\phi^{\prime}_L}{2\pi}
\label{charge0La2}.
\ee
Reverting back to the original boundary variables, we have

\be\label{charge0La2or}\mathcal{N}=\frac{\phi_R+\phi_L}{2\pi}.\ee
Absorbing the boundary fields into the normal ordering as before (\ref{Nprime}), we have

\be \mathcal{N}=0.
\ee
Hence just as before, we label this state $\ket{0}_L$. Now consider adding the boundary string $\pm i (u+\phi^{\prime}_L)$ corresponding to the left boundary to the state $\ket{1}$. Labelling the density distribution of the resulting state as $\rho^{\prime}_{\phi_L}(\alpha)$, we have

\bea \nonumber2m_0\cosh\alpha-\frac{1}{L}\varphi^{\prime}(\alpha,-\phi_R+u)-\frac{1}{L}\varphi^{\prime}(\alpha,\phi^{\prime}_L+u)-4\pi\rho^{\prime}_{\phi_L}(\alpha)-\frac{1}{L}\delta(\alpha)+\frac{1}{L}\varphi^{\prime}(\alpha,u)+\frac{1}{L}\varphi^{\prime}(\alpha,u+i\pi)\\=\frac{1}{L}\varphi^{\prime}(\alpha,\phi_L+3u)+\frac{1}{L}\varphi^{\prime}(\alpha,u-\phi_L)+\sum_{\sigma}\int\varphi^{\prime}(\alpha+\sigma\gamma,2u)\rho^{\prime}_{\phi_L}(\gamma)d\gamma.
\label{gsdenbsa2l}\eea
Following the same procedure as above, we obtain  

\be \tilde{\rho'}_{\phi_L}(\omega)=\tilde{\rho}_{\ket{1}}(\omega)+\Delta\tilde{\rho'}_{\phi_L}(\omega), \;\;\; \Delta\tilde{\rho'}_{\phi_L}(\omega)=-\frac{1}{L}\;\frac{\sinh((\pi-2u)\omega)\cosh((u+\phi^{\prime}_L)\omega)}{\sinh(\pi\omega)+\sinh((\pi-2u)\omega)},\ee
where $\tilde{\rho}_{\ket{1}}(\omega)$ is given by (\ref{denforma2+1}), (\ref{denba2+1}). The total charge of this state is

\be 1+L\tilde{\rho'}_{\phi_L}(0)=\frac{m_0 L e^{\Lambda}}{2\pi}\;\frac{\pi}{2(\pi-u)}-\frac{\phi_R-\phi^{\prime}_L}{4(\pi-u)}.\ee
Just as before, total charge of the above state is less than the maximum allowed charge (\ref{MaxC}), and hence this is a valid state. The normal ordered charge of this state is 
\be
\mathcal{N}=\frac{\phi_R-\phi^{\prime}_L}{2\pi}
\label{charge0Ra2}.
\ee
Reverting back to the original boundary variables, we have

\be\label{charge0Ra2or}\mathcal{N}=\frac{\phi_R+\phi_L}{2\pi}.\ee
Absorbing the boundary fields into the normal ordering just as before (\ref{Nprime}), we have

\be\mathcal{N}=0.
\ee
Hence we label this state $\ket{0}_R$. Let us denote the energies of the boundary strings $\pm i(u-\phi_R)$ and $\pm i (u+\phi^{\prime}_L)$ by $E_{\phi_R}$ and $E_{{\phi}^{\prime}_L}$ respectively. These energies can be calculated in the same way as in the $A_1$ phase. We obtain

\be E_{\phi_R}=-m_R=-m\sin\gamma\phi_R, \;\;\; E_{{\phi}^{\prime}_L}=m'_L=m\sin\gamma\phi^{\prime}_L=-m_L=-m\sin\gamma\phi_L.
\ee
The energy difference between the states $\ket{1}$ and $\ket{-1}$ can be calculated in a similar way as in the $A_1$ phase. We obtain

\be E_{\ket{1}}-E_{\ket{-1}}=m_R+m_L.
\ee
Since $\phi_L<0,\phi_R>|\phi_L|$, we have $m_L<0,m_R>|m_L|$. Hence, the energy of the state $\ket{-1}$ which is $E_{\ket{1}}-(m_R+m_L)$, is lower than that of the state $\ket{1}$. By adding the right boundary string to the state $\ket{1}$ we obtain the state $\ket{0}_L$, which has energy $E_{\ket{1}}-m_R$. Similarly, by adding the left boundary string to the state $\ket{1}$ we obtain the state $\ket{0}_R$, which has energy $E_{\ket{1}}-m_L$. Comparing the energies of these four states, we see that the state $\ket{0}_L$ is the ground state in the $A_2$ phase for $\phi_R>|\phi_L|$.

\subsubsection{The states $\ket{\pm 1} ( \phi_R<|\phi_L|$) }
Now let us consider the case of  $\phi_R<|\phi_L|$. In this region, the Bethe equations are exactly the same as in the previous case (\ref{BEsupA2+1}). The states $\pm 1$ are constructed in the exact same way and they are described by the same density distributions obtained in the previous case. The charge of the state $\ket{1}$ is given by (\ref{charge+1A2}), which can be written as

\be \label{charge+1A22} L\tilde{\rho}_{\ket{1}}(0)=\frac{m_0 L e^{\Lambda}}{2\pi}\;\frac{\pi}{2(\pi-u)}-\frac{\pi}{2(\pi-u)}+\frac{\phi^{\prime}_L-\phi_R}{4(\pi-u)}.
\ee
Similarly, the charge of the state $\ket{-1}$ is given by (\ref{charge-1A2}), which is

\be \label{charge-1A22} L\tilde{\rho}_{\ket{-1}}(0)=\frac{m_0 L e^{\Lambda}}{2\pi}\;\frac{\pi}{2(\pi-u)}-\frac{\pi}{2(\pi-u)}-\frac{\phi^{\prime}_L-\phi_R}{4(\pi-u)}.
\ee
Notice that the first two terms are exactly the same as in the previous case (\ref{charge-1A2}),(\ref{charge+1A2}), but the sign of the last term is different.  Due to this, the construction of the states $\ket{0}_{\cal L}$ and $\ket{0}_{\cal R}$ is different compared to the previous case, which we discuss below.

\subsubsection{The states $\ket{0}_{\cal L}$ and $\ket{0}_{\cal R}$ $(\phi_R<|\phi_L|)$}
From the above expressions for the charges of the states $\ket{\pm 1}$, we see that unlike in the previous case $\phi_R>|\phi_L|$, the difference between the charge of the state $\ket{-1}$ and the maximum allowed charge (\ref{MaxC}) is large enough in order for a close (short) boundary string to be added to it. The Bethe equations corresponding to the original fermions in the $A_2$ phase have two boundary string solutions: $\pm i (u+\phi_R)$, $\pm i (u-\phi^{\prime}_L)$. Consider adding the right boundary string $\pm i (u+\phi_R)$ to the state $\ket{-1}$. Following the same procedure as before, we obtain

\be \tilde{\rho'}_{\phi_R}(\omega)=\tilde{\rho}_{\ket{-1}}(\omega)+\Delta\tilde{\rho'}_{\phi_R}(\omega), \;\;\; \Delta\tilde{\rho'}_{\phi_R}(\omega)=-\frac{1}{L}\;\frac{\sinh((\pi-2u)\omega)\cosh((u+\phi_R)\omega)}{\sinh(\pi\omega)+\sinh((\pi-2u)\omega)},\ee
where $\tilde{\rho}_{\ket{-1}}(\omega)$ is given by (\ref{dena2-1}),(\ref{denba2-1}). The total charge of this state is

\be 1+L\tilde{\rho'}_{\phi_R}(0)=\frac{m_0 L e^{\Lambda}}{2\pi}\;\frac{\pi}{2(\pi-u)}-\frac{\phi^{\prime}_L-\phi_R}{4(\pi-u)}.\ee
Since $\phi_R< \phi^{\prime}_L (|\phi_L|)$, the total charge of the above state is less than the maximum allowed charge (\ref{MaxC}), and hence this is a valid state. The normal ordered charge of this state is 
\be
\mathcal{N}=-\frac{\phi^{\prime}_L-\phi_R}{2\pi}
\label{charge0RA22}.
\ee
Reverting back to the original boundary variables, we have

\be\label{charge0RA22or}\mathcal{N}=\frac{\phi_R+\phi_L}{2\pi}.\ee
Just as before (\ref{Nprime}), by absorbing the boundary fields into the normal ordered charge we have

\be\mathcal{N}=0.
\ee
Now consider adding the boundary string $\pm i (u-\phi^{\prime}_L)$ corresponding to the left boundary to the state $\ket{-1}$. Following the same procedure as above, we obtain  

\be \tilde{\rho}_{\phi_L}(\omega)=\tilde{\rho}_{\ket{-1}}(\omega)+\Delta\tilde{\rho}_{\phi_L}(\omega), \;\;\; \Delta\tilde{\rho}_{\phi_L}(\omega)=-\frac{1}{L}\;\frac{\sinh((\pi-2u)\omega)\cosh((u-\phi^{\prime}_L)\omega)}{\sinh(\pi\omega)+\sinh((\pi-2u)\omega)},\ee
where $\tilde{\rho}_{\ket{-1}}(\omega)$ is given by (\ref{dena2-1}),(\ref{denba2-1}). The total charge of this state is

\be 1+L\tilde{\rho}_{\phi_L}(0)=\frac{m_0 L e^{\Lambda}}{2\pi}\;\frac{\pi}{2(\pi-u)}-\frac{\phi^{\prime}_L-\phi_R}{4(\pi-u)}.\ee
Just as before, the total charge of the above state is less than the maximum allowed charge (\ref{MaxC}), and hence this is a valid state. The normal ordered charge of this state is 
\be
\mathcal{N}=-\frac{\phi^{\prime}_L-\phi_R}{2\pi}
\label{charge0LA22}.
\ee
Reverting back to the original boundary variables, we have

\be\label{charge0LA22or}\mathcal{N}=\frac{\phi_R+\phi_L}{2\pi}.\ee
Just as before (\ref{Nprime}), by absorbing the boundary fields into the normal ordered charge we have

\be\mathcal{N}=0.
\ee
Let us denote the energies of the boundary strings $\pm i(u+\phi_R)$ and $\pm i (u-\phi^{\prime}_L)$ by $E_{{\phi}^{\prime}_R}$ and $E_{\phi_L}$ respectively. These energies can be calculated in the same way as before. We obtain

\be E_{{\phi}^{\prime}_R}=m_R, \;\;\; E_{\phi_L}=m_L.
\ee
Just as before, the energy difference between the states $\ket{1}$ and $\ket{-1}$ is

\be E_{\ket{1}}-E_{\ket{-1}}= m_R+m_L.
\ee
Since $\phi_L<0,\phi_R<|\phi_L|$, we have $m_L<0,|m_L|>m_R$. Hence, the energy of the state $\ket{1}$ which is $E_{\ket{-1}}+(m_L+m_R)$, is lower than that of the state $\ket{-1}$. By adding the right boundary string to the state $\ket{-1}$ we obtain the state which has energy $E_{\ket{-1}}+m_R$. This is precisely the state $\ket{0}_{\cal R}$ with bound state at the right boundary. Similarly, by adding the left boundary string to the state $\ket{-1}$ we obtain the state $\ket{0}_{\cal L}$ which hosts bound state at the left boundary
and has energy $E_{\ket{-1}}+m_L$. Comparing the energies of these four states, we see that the state $\ket{0}_L$ is the ground state in the $A_2$ phase for $\phi_R<|\phi_L|$ as well. Comparing the energies of the states $\ket{\pm1}$ in the two regions ($\phi_R<|\phi_L|$) and ($\phi_R>|\phi_L|$), we find that they are degenerate on the line $\phi_R=-\phi_L$.

 \subsubsection{Summary Phase $A_2$:} 
  
  1) Just as in the $A_1$ phase, there exists four states $\ket{\pm1}, \ket{0}_{L}, \ket{0}_{R}$ whose energy is below the mass gap $m$. 
  
  2) Both the states $\ket{\pm1}$ contain Bethe roots which lie on the $i\pi$ line. The energy difference between these states is $E_{\ket{1}}-E_{\ket{-1}}=m_R+m_L$, where $m_R=m\sin\gamma\phi_R$, $m_L=m\sin\gamma\phi_L$.
  
3) Comparing the energies of the two states $\ket{1}$ and $\ket{-1}$, we find that they are degenerate on the line $\phi_L=-\phi_R$. 

4) For $\phi_R>|\phi_L|$, the states $\ket{0}_{R},\ket{0}_{L}$ are obtained by adding the boundary strings $\pm i(u+\phi^{\prime}_{L}), \pm i(u-\phi_R)$ to the state $\ket{1}$ respectively, where $\phi^{\prime}_L\equiv |\phi_L|$. For  $\phi_R<|\phi_L|$, the states $\ket{0}_{R},\ket{0}_{L}$ are obtained by adding the boundary strings $\pm i(u+\phi_{R}), \pm i(u-\phi^{\prime}_L)$ to the state $\ket{-1}$ respectively. 

5) The boundary strings $\pm i(u-\phi_{R})$, $\pm i(u-\phi^{\prime}_L)$ have energies $-m_{R},m_L$ respectively, and 
the boundary strings $\pm i (u+\phi_{R})$, $\pm i(u+\phi^{\prime}_L)$ have energies $m_{R}, -m_L$ respectively. Since $\phi_L<0, \phi_R>|\phi_L|$, we have $m_L<0, m_R>|m_L|$. Comparing the energies of the four states $\ket{\pm1}$,$\ket{0}_{\cal L,R}$ we see that the ground state in the $A_2$ phase is $\ket{0}_L$.

\subsection{$A_3$ and $A_4$ phases}
The Bethe equations in the $A_3$ and $A_4$ phases can be obtained by making the transformation $\phi_{L,R}\rightarrow -\phi_{L,R}$ corresponding to the Bethe equations of the $A_1$ and $A_3$ phases respectively. Hence, the Bethe equations corresponding to the original (charge conjugated) fermions of $A_3$ and $A_4$ phases are exactly the Bethe equations corresponding to the charge conjugated (original) fermions of $A_1$ and $A_2$ phases respectively. The above analysis of the $A_1$ and $A_2$ phases can be repeated in the exact same way to obtain the results in the $A_3$ and $A_4$ phases respectively. The results are summarized below.

\subsubsection{$A_3$ phase}

In the $A_3$ phase, the ground state belongs to the odd fermionic parity sector and it is $\ket{+1}$. The state $\ket{-1}$ has energy $m_L+m_R$ above the ground state. The energies of the states $\ket{0}_{\cal{L}}$ and $\ket{0}_{\cal{R}}$ with respect to the ground state are $m_R$ and $m_L$ respectively. Hence, in the even fermionic parity sector, just as in the case of $A_1$ phase, the lowest energy state depends on the values of the boundary fields $\phi_{L,R}$: For $|\phi_R|<|\phi_L|$, the lowest energy state is $\ket{0}_{\cal{L}}$, whereas for $|\phi_R|>|\phi_L|$, it is $\ket{0}_{\cal{R}}$. On the line $\phi_L=\phi_R$, like in the $A_1$ phase, the lowest energy state is two fold degenerate due to the space parity symmetry $\mathbb{P}$.

\subsubsection{$A_4$ phase}

In the $A_4$ phase, the ground state belongs to the even fermionic parity sector, and it is $\ket{0}_{\cal{R}}$. The state $\ket{0}_{\cal{L}}$ has energy $m_L+m_R$ above the ground state $\ket{0}_{\cal{R}}$. The energies of the states $\ket{-1}$ and $\ket{1}$ with respect to the ground state are $m_R$ and $m_L$ respectively. Hence, in the odd fermionic parity sector, just as in the $A_2$ phase, the state with the lowest energy depends on the values of $\phi_{L,R}$: For $\phi_L>|\phi_R|$ the lowest energy state is $\ket{-1}$, whereas for $\phi_L<|\phi_R|$ it is $\ket{+1}$. Just as in the $A_2$ phase, there exist two lowest energy states  with same energy on the line $\phi_R+\phi_L=0$, which occurs due to the system being invariant under the combined space parity $\mathbb{P}$ and charge conjugation $\mathcal{C}$ transformations.

\end{widetext}

\end{document}